\documentclass[aps,superscriptaddress,showpacs,amssymb,amsmath,prb,11pt]{revtex4-1}

\usepackage{graphicx}
\usepackage[pdftex, colorlinks=true, linkcolor=blue, citecolor=blue, urlcolor=blue]{hyperref}

\usepackage{units}
\usepackage{txfonts}
\usepackage{bm}
\usepackage{soul}

\newcommand{\z}{\text {z}}
\newcommand{\dint}{\mathrm{d}}

\begin{document}
\title{Spin-relaxation time in the impurity band of wurtzite semiconductors}

\author{Pablo I.\ Tamborenea}
\affiliation{Departamento de F\'{\i}sica and IFIBA, FCEN, Universidad de Buenos Aires, 
Ciudad Universitaria, Pab.\ I, C1428EHA Ciudad de Buenos Aires, Argentina}

\author{Thomas Wellens}
\affiliation{Physikalisches Institut der Albert-Ludwigs-Universit\"{a}t, 
Hermann-Herder-Str. 3, D-79104 Freiburg, Germany}

\author{Dietmar Weinmann}
\author{Rodolfo A.\ Jalabert}
\affiliation{Universit\'e de Strasbourg, CNRS, Institut de Physique et Chimie des Mat\'{e}riaux 
de Strasbourg, UMR 7504, F-67000 Strasbourg, France}

\begin{abstract}
The spin-relaxation time for electrons in the impurity band of semiconductors
with wurtzite crystal structure is determined.
The effective Dresselhaus spin-orbit interaction Hamiltonian is taken as the source of
the spin relaxation at low temperature and for doping densities corresponding to the 
metallic side of the metal-insulator transition.
The spin-flip hopping matrix elements between impurity states are calculated and used to set up 
a tight-binding Hamiltonian that incorporates the symmetries of
wurtzite semiconductors.
The spin-relaxation time is obtained from a semiclassical model of spin diffusion,
as well as from a microscopic self-consistent diagrammatic theory of spin 
and charge difussion in doped semiconductors.
Estimates are provided for particularly important materials.
The theoretical spin-relaxation times compare favorably with the corresponding
low-temperature measurements in GaN and ZnO.
%
For InN and AlN, we predict that tuning of the spin-orbit coupling constant induced 
by an external potential leads to a potentially dramatic increase of the spin-relaxation time 
related to the mechanism under study.
\end{abstract}

\pacs{72.25.Rb, 76.30.Pk, 72.20.Ee, 03.67.-a}
\maketitle

\section{Introduction}

The group-III nitride and the group-II oxide semiconductors have direct bandgaps, 
which cover the ultraviolet to infrared energy range.\cite{Hanada2009}
In particular, the wide bandgaps of AlN, GaN, and ZnO make these materials extremely 
important for optoelectronic and high-power applications. 
In the case of ZnO, the present interest is also fueled by the large exciton binding energy 
that allows for high efficiency light emission up to high temperatures, 
by the prediction that manganese doping induces room-temperature 
ferromagnetism,\cite{Dietl1019,pearton-JAP2003} as well as by the relatively simple 
crystal-growth technology involved.\cite{ozgur-JAP2005,janotti-RPP2009} 
A key characteristic of this class of materials, linked with very special physical properties, 
is that their stable crystalline structure is of hexagonal wurtzite (WZ) type. 
The light nuclei result in a small valence band splitting of otherwise degenerate spin states, 
leading to a relatively weak spin-orbit coupling, based on which long spin coherence times
have been anticipated. \cite{kri-van-new}
A long spin lifetime is a necessary ingredient for spintronics and quantum technology 
applications,\cite{sou-rey-fer-pan,aws-bas-dzu,bad-par,fab-mat-ert} paving the way to 
all-semiconductor spintronic devices based on this group of 
materials.\cite{awschalom-book2002,zut-fab-das}

Despite the interesting physical properties and the technological promise of 
the group-III nitride and the group-II oxide semiconductors (and their alloys) 
with WZ structure, the corresponding spin properties have been less extensively studied, 
at the experimental and theoretical levels, than in the case of III-V cubic 
zincblende (ZB) semiconductors. 
In the latter context, detailed measurements have been performed for n-doped GaAs, yielding 
very long spin lifetimes (of the order of hundreds of nanoseconds) at 
low temperatures.\cite{kik-aws,dzh-kav-kor,oestreich-PRL2005} 
In particular, density-dependent measurements yielded the longest spin-relaxation times 
for doping concentrations in the neighborhood of the critical one for the metal-insulator 
transition (MIT). 
This intriguing behavior has contributed to the sustained theoretical interest 
that ZB materials have received.\cite{wu-jia-wen}
In the regime of longest spin-relaxation times, for doping densities between the critical one 
and that where the impurity band hybridizes with the conduction band, a theoretical description 
based in mechanisms relating spin relaxation to momentum scattering, like Dyakonov-Perel and 
Elliot-Yafet, cannot be applied from a conceptual point of view. 
This regime has been addressed, at low temperature and in the absence of magnetic field, 
in terms of a tight-binding Hamiltonian \cite{tamborenea-PRB2007} which incorporates spin-orbit 
coupling into the Matsubara-Toyozawa model.\cite{matsubara-1961} 
The inclusion of the bulk Dresselhaus spin-orbit interaction 
resulted in a good agreement with the observed spin lifetimes 
for several ZB materials. \cite{intronati-PRL2012, wellens-jalabert-2016}

The existing low-temperature spin-relaxation experiments in WZ semiconductors 
point to a scenario that is similar to the ZB case. 
For Si-doped GaN, the dependence of the relaxation time on both magnetic field 
and temperature was found to be qualitatively similar to previous studies in 
n-type GaAs, suggesting a common origin for spin relaxation in these 
systems.\cite{bes-etal}
Indeed, at $T=\unit[5]{K}$, out of three studied samples with different doping 
densities, the one having a doping density in the vicinity of the MIT yielded 
the longest relaxation time (of about \unit[20]{ns}).
These values of the spin-relaxation time were confirmed in the framework of 
a detailed study of the properties of the MIT in GaN.\cite{wol-etal}
The spin-relaxation time was found to exhibit a temperature-dependent maximum
as a function of doping density.\cite{bus-etal-2010a,bus-etal-2011}
In the temperature interval of these studies ($T=\unit[80]{K}$ to room temperature)
the electronic conductance is dominated by conduction-band properties and therefore
the spin relaxation is consistent with the Dyakonov-Perel mechanism.

Spin-relaxation times of up to 20 ns have been measured in doped ZnO at 30 K.
Bulk and epilayer samples of different doping densities were investigated
and the longest relaxation time was found for the bulk sample.\cite{gho-etal}
The temperature dependence of the relaxation rate was found to be consistent 
with the Dyakonov-Perel mechanism.\cite{pre-etal} 
It was noticed that a static in-plane electric field 
can enhance the spin lifetime in a quantum well geometry.\cite{ghosh-APL2008}

Aside from Refs.\ [\onlinecite{bes-etal}], [\onlinecite{wol-etal}], 
and [\onlinecite{gho-etal}], where samples with different impurity 
concentrations have been measured, there is no systematic 
study of the dependence on doping density of the low-temperature spin lifetime 
close to the metal-insulator transition for the WZ materials. 
Most experimental data are not for the low-temperature regime, 
which is where the impurity-band physics is dominant and where the longest 
lifetimes are expected. 

From the theoretical point of view, the ZB and the WZ crystal structures have 
the same tetrahedral nearest-neighbor atomic coordination 
number
and share 
the lack of inversion symmetry that results in the splitting of spin-degenerate states. 
The main qualitative difference between them is that the latter presents 
a uniaxial anisotropy which is absent in the former. 
At the level of the envelope-function approximation this anisotropy affects 
the Dresselhaus spin-orbit effective interaction in the conduction band: 
its cubic-in-$k$ term becomes anisotropic and a linear-in-$k$ term 
(formally identical to the Rashba interaction in quantum wells) appears. 
The linear term has an intrinsic contribution that may depend on 
substrate-dependent strain in epilayers, and an external one that 
can be controlled by applying electric fields. 
The resulting spin relaxation emerges from an interplay of the effects of the cubic 
and the linear spin-orbit couplings, and can then be influenced by tuning the 
linear coupling strength.\cite{ganichev-pss2014,harmon-APL2011}

An important quantitative difference between the WZ and the ZB structures is 
the much smaller valence-band splitting of the former as compared with the latter. 
Such feature would generically point to a much larger spin-relaxation time for 
WZ structures than in the ZB case. 
The existing experiments do not validate this simplistic conclusion, 
indicating the important influence on the spin-relaxation mechanisms 
of the uniaxial anisotropy as well as other physical parameters.  

Spin relaxation in the conduction band of WZ structures has been theoretically addressed, 
using the Dyakonov-Perel formalism,\cite{harmon-PRB2009,harmon-APL2011,wang-JAP2010} 
and predicting particularly long spin lifetimes for the case of AlN, 
for a narrow range of values of the linear coupling. 
The temperature- and magnetic field dependence of the spin lifetime was investigated 
in Ref.\ [\onlinecite{kang-ps2015}].
The anisotropy of the spin-orbit coupling was found to yield a dependence 
of the Dyakonov-Perel spin-relaxation rate on the initial 
orientation,\cite{harmon-PRB2009,wang-JAP2010,bus-etal-2009} 
with a decay twice as fast for the component along the crystal symmetry axis 
as compared to in-plane spin.
Such an asymmetry has also been obtained in the context of ZB quantum wells with particular 
crystal orientations. \cite{cartoixa-PRB2005}
The experimentally observed \cite{gho-etal} nonmonotonic behavior of the spin lifetime
as a function of temperature in ZnO has been explained invoking the spin exchange between 
localized and extended states.\cite{harmon-PRB2009}
Alternatively, the numerical solution of Bloch equations for n-type ZnO quantum wells
yielded a maximum relaxation time as a function of temperature, and also as a function
of carrier density for sufficiently low temperatures. \cite{lu-SST2009}
These theoretical studies rely on material parameters and spin-orbit coupling constants that 
have been extracted from numerical band-structure 
calculations,\cite{fu-wu, wang-etal,wang-JAP2010, xu-PRB1993}
and therefore introduce a certain level of uncertainty in the theoretical predictions.

It is important to remark that the theoretical methods used so far to study WZ structures 
are not expected to be applicable to the low-temperature physics on the metallic side 
of the metal-insulator transition in the impurity band. \cite{matsubara-1961,tamborenea-PRB2007}
The rather incomplete experimental and theoretical information on spin physics 
for the impurity band of WZ semiconductors calls for further studies to explore 
whether long spin-relaxation times are possible in the vicinity of the metal-insulator 
transition in the impurity band of n-doped semiconductors having WZ crystal structure.

In this article we adapt theoretical methods developed in Refs.\ [\onlinecite{intronati-PRL2012}] 
and [\onlinecite{wellens-jalabert-2016}] in the context of ZB semiconductors, 
in order to calculate the spin-relaxation time in the metallic side 
of the impurity band of WZ bulk semiconductors, 
and compare our results with the experimental data available in the literature
for GaN and ZnO.
We also study how far the spin-relaxation time can in principle be extended by tuning 
the linear-in-$k$ component of the Dresselhaus spin-orbit interaction.
We find that this tuning is promising for GaN and potentially dramatic for AlN.

From a practical point of view, the understanding of spin-relaxation mechanisms 
in WZ bulk semiconductors serves a two-fold purpose. 
One one hand, important information can be obtained in the cases where materials 
having ZB crystal structure in the bulk turn to a WZ configuration when nanostructured 
in nanorods or quantum 
dots. \cite{de-pry,koguchi-JJAP92,intronati-PRB2013,intronati_thesis}
On the other hand, even if the spin relaxation can be considerably slowed down 
in semiconductor heterostructures and nanostructures, doped bulk semiconductors 
are always required for the contact layers of the devices.
From the fundamental point of view, it is important to understand 
how the symmetries of the WZ structure affect the spin-relaxation mechanisms, 
and also to have available a tool to better characterize the 
metal-insulator transition in wide bandgap semiconductors.

This article is organized as follows.
In Section \ref{sec:hamiltonian} we introduce the Hamiltonian and other basic elements 
of the tight-binding model with Dresselhaus spin-orbit interaction used to describe 
an electron in the impurity band of bulk wurtzite semiconductors.
In particular, we obtain the hopping matrix elements of the Dresselhaus spin-orbit terms 
of the Hamiltonian; details of this calculation are given in Appendix \ref{sec:appendixA}.
Sections \ref{sec:semiclassical} and \ref{sec:ssa} present, respectively, 
the semiclassical and the fully quantum-mechanical self-consistent approach 
to the spin-relaxation time. 
The latter with three successive degrees of approximation. 
In Sec.\ \ref{sec:qr} we apply our theory to four specific semiconductors and examine 
the proposed scheme for maximization of the spin lifetime.
Our conclusions are given in Section \ref{sec:conclusion}.
For completeness, in Appendix \ref{sec:appendixB} we apply the semiclassical formula 
to the case of zincblende materials.
Relevant Fourier transforms are given in Appendix \ref{app:ftham}.

\section{Hamiltonian and hopping amplitude matrix}
\label{sec:hamiltonian}

The envelope-function approximation for conduction-band electrons in WZ semiconductors 
incorporates the crystal lattice-scale physics into
the effective one-body Hamiltonian \cite{noz-lew,eng-ras-hal}
\begin{subequations}
\label{eq:Hallunrestr}
\begin{eqnarray} 
H &=& H_0 + H_{\text{D}} + H_{\text{extr}} \, ,
\label{eq:Htot}
\\ 
H_0 &=& \frac{p^2}{2 \, m^*} + V(\mathbf{r}) \, ,
\label{eq:Hzero}
\\ 
H_{\text{D}} &=& H_{\text{D},\text{1}} + H_{\text{3}}  \, , 
\label{eq:BIA}
\\
H_{\text{extr}} &=& \lambda \, \boldsymbol{\sigma} \cdot \nabla V 
                 \times \mathbf{k} \, .
\label{eq:SIA}
\end{eqnarray}
\end{subequations}
The spin-independent part $H_0$ is determined by the effective mass ($m^*$) and 
the electrostatic potential $V(\mathbf{r})$ including all potentials aside 
from that of the crystal lattice. 
We note $\mathbf{p}$ the momentum operator, 
$\mathbf{k} = \mathbf{p}/\hbar$, and $\boldsymbol{\sigma}$ the vector of Pauli matrices. 
The Dresselhaus (intrinsic) term $H_{\text{D}}$, enabled by the bulk inversion asymmetry, 
comprises in the case of WZ semiconductors, the linear- \cite{lew-wil-car-chr} and 
the cubic-in-$k$ \cite{wang-etal,fu-wu} components, respectively given by 
\begin{subequations}
\label{eq:HDress}
\begin{eqnarray} 
H_{\text{D},\text{1}} &=& \alpha_{\mathrm{D}} \, (k_y \sigma_x - k_x \sigma_y) \, , 
\label{eq:linear} \\
H_{\text{3}} &=& \gamma \, (b k_z^2 - k_x^2 - k_y^2) 
                           (k_y \sigma_x - k_x \sigma_y) \, ,
\label{eq:cubic}
\end{eqnarray}
\end{subequations}
and therefore depends on three material-dependent constants, namely, $\alpha_\mathrm{D}$, $\gamma$, 
and $b$. 
These parameters are in some cases only partially known, as will be discussed in our treatment 
of specific materials in Sec.~\ref{sec:qr}. 
The extrinsic term $H_{\text{extr}}$ of Eq.~\eqref{eq:SIA} has the same form as the spin-orbit 
interaction in vacuum, but with a material-dependent renormalized coupling constant $\lambda$.
External potentials translating into an electric field in $z$-direction result in a Rashba 
contribution to $H_{\text{extr}}$ that has the same linear-in-$k$ form of  $H_{\text{D},\text{1}}$, 
with a prefactor $\alpha_\mathrm{R}$ that depends on $\lambda$. 
While such a field cannot be significant in the bulk case, perpendicular electric fields 
in thin films might play a role. 
In the latter case, the strain induced on the film by the substrate could be 
an additional mechanism to modify the linear-in-$k$ coupling.
The linear-in-$k$ contributions can be jointly treated by defining a Hamiltonian $H_1$ given 
by \eqref{eq:linear}, but with a coupling coefficient
\begin{equation}
\alpha = \alpha_\mathrm{D}+\alpha_\mathrm{R} \, .
\label{eq:alphas}
\end{equation}
Separating between Dresselhaus and Rashba contributions with the same functional form 
is generically a challenge, that has been experimentally addressed in the case 
of ZB quantum wells with the application of an external magnetic 
field.\cite{meier-NatPhy2007} 
Given the uncertainties in the material parameters, we will not attempt 
such a separation in our theoretical treatment of the WZ structure and 
we will express our results as a function of the tunable parameter $\alpha$.
In the literature, $H_1$ is often simply referred to as the "Rashba" 
Hamiltonian \cite{fu-wu,ste-2014}, 
while the cubic-in-$k$ contribution, $H_3$, is labeled as the "Dresselhaus" term. 
We will avoid here this somewhat misleading nomenclature.

The potential arising from the ionized donor impurities is given by
\begin{equation} 
V_{\text{imp}}(\mathbf{r}) = \sum_m V_m(\mathbf{r}) = 
                - \sum_m \frac{e^2}{\epsilon |\mathbf{r}-\mathbf{R}_m|} \, ,
\end{equation}
where $\epsilon$ is the dielectric constant of the semiconductor and 
$\mathbf{R}_m$ represents the impurity positions. 
Its contribution to $H_{\text{extr}}$ has been shown to be extremely weak in the case 
of ZB structures, \cite{tamborenea-PRB2007} and we expect the same considerations to 
also hold in the WZ case. 
Therefore, it will not be considered in our theoretical analysis. 
On the other hand, the potential $V_{\text{imp}}(\mathbf{r})$ strongly affects the orbital 
motion of the electrons, and will be the only contribution to $V(\mathbf{r})$ that we 
will keep in Eq.~\eqref{eq:Hzero}.

The potential $V_m(\mathbf{r})$ gives rise to hydrogenic states centered 
at the impurity $m$, with a ground-state wave-function 
$\phi_m(\mathbf{r})= \phi(|\mathbf{r}-\mathbf{R}_m|)$.
The anisotropy of the WZ crystal lattice induces a uniaxial anisotropy, 
of about ten percent, in the effective masses and the dielectric 
constants. \cite{Hanada2009,per-fer-ahu-joh, fer-ara} 
We will neglect the resulting small deformation of the hydrogenic impurity wave 
functions, \cite{Rodina2001} adopting the standard isotropic ground-state 
wave-function $\phi(\mathbf{r}) = (1/\sqrt{\pi a^3}) \exp{(-r / a)}$, 
where $a$ stands for the effective Bohr radius. 
The electronic ground states of the different sites $m$ provide a restricted basis 
$\{| m \sigma \rangle\}$ to describe the electron jumping between impurity centers 
($\sigma=\pm$ corresponds to a spin projection of $\pm \hbar/2$ in the $z$-direction). 
Choosing as energy origin the ground-state energy, the Hamiltonian in this restricted 
space can be expressed as 
\begin{equation}
 {\mathcal H} = \sum_{m'\neq m} \sum_{\sigma',\sigma} 
                            |m'\sigma'\rangle \,
                            \mathcal{V}^{\sigma',\sigma}(\mathbf{R}_{m'm}) \, 
                            \langle m \sigma| \, ,
\end{equation}
where $\mathbf{R}_{m'm} = \mathbf{R}_{m'} - \mathbf{R}_m$. 
In the following, we neglect the overlaps between different states $m \neq m'$ 
and thereby treat $\{|m\sigma\rangle\}$ as an orthonormal basis, 
which is justified in the regime that we are interested in 
(i.e., for moderate doping densities and energies close to the center of the impurity band)
\cite{maj-and_83,fig-mak-and_84}.
Like in the ZB case, the hopping amplitudes can be generically expressed through a $2 \times 2$ 
matrix in the spin subspace\citep{wellens-jalabert-2016}
\begin{equation}\label{eq:spin_matrix_element}
\mathcal{V}(\mathbf{r})=\left(\begin{array}{cc}
          \mathcal{V}_0(\mathbf{r})+i\mathcal{C}_z(\mathbf{r})&
                     i\mathcal{C}_x(\mathbf{r})+\mathcal{C}_y(\mathbf{r})\\
          i\mathcal{C}_x(\mathbf{r})-\mathcal{C}_y(\mathbf{r}) 
          &\mathcal{V}_0(\mathbf{r})-i\mathcal{C}_z(\mathbf{r})
                            \end{array}\right).
\end{equation}
The spin-independent hopping amplitudes are those of the Matsubara-Toyozawa model \cite{matsubara-1961} 
\begin{equation} 
{\mathcal V}_{0}({\bf r}) =  -V_0 \left(1+\frac{r}{a} \right) \ e^{-r/a} 
\, .
\label{eq:meh0}
\end{equation}
We note ${\bf r}=(x,y,z)$ and $r=|{\bf r}|$, while $V_0= 
2 E_\mathrm{Ry}^{(0)}m^*/\epsilon^2$ corresponds to twice the binding energy of the 
impurity sites ($E_\mathrm{Ry}^{(0)} = \unit[13.6]{meV}$ is the Rydberg energy).

The matrix elements of the linear-in-$k$ spin-orbit Hamiltonian, $H_1$, are
\begin{eqnarray} 
\langle m'\sigma'|H_1|m\sigma\rangle &=& 
  \alpha \langle m'\sigma' | k_y \sigma_x - k_x \sigma_y | m\sigma\rangle 
  \nonumber \\
  &=& \alpha \delta_{\sigma',\overline{\sigma}}
  \left(\langle m'|k_y|m\rangle - i\sigma \langle m'|k_x|m\rangle \right) \, .
\end{eqnarray}
We use the notation $\overline{\sigma}=-\sigma$, and thus
$\sigma_x |\sigma \rangle= |\overline{\sigma}\rangle$
and
$\sigma_y |\sigma \rangle=i\sigma |\overline{\sigma}\rangle$.
Applying the operators $k_x=-i \partial_x$ and $k_y=-i \partial_y$ we obtain
\begin{equation} 
\label{eq:H1b}
\langle m'\sigma'|H_1|m\sigma\rangle = 
   \frac{\sigma \alpha}{a} \, 
   \delta_{\sigma',\overline{\sigma}} \,
   \int \dint \mathbf{r} \, \phi_{m'}^*(\mathbf{r}) 
               \frac{(x-X_m)+i\sigma (y-Y_m)}{|\mathbf{r}-\mathbf{R}_m|} 
             \phi_m(\mathbf{r}) \, .
\end{equation}
In Appendix \ref{sec:appendixA} we perform the analytical integration which results in
\begin{equation} \label{eq:res_matrix_element_1}
\langle m'\sigma'|H_1|m\sigma\rangle =  
   \frac{\sigma \alpha}{3a^2} \, 
   \delta_{\sigma' \overline{\sigma}} \,
    R_{m'm} \sin\theta \, e^{i\sigma\phi}  
    \, \left(1 + \frac{R_{m'm}}{a} \right)
    \, e^{-R_{m'm}/a} \, ,
\end{equation}
where
$\mathbf{R}_{m'm} = 
R_{m'm} (\sin\theta \cos\phi \, \mathbf{x} + \sin\theta \sin\phi \, 
\mathbf{y} + \cos\theta \, \mathbf{z})$.



\bigskip

As shown in Appendix \ref{sec:appendixA}, the matrix elements of the cubic-in-$k$ term 
\begin{equation}
\label{eq:res_matrix_element_3f}
\langle m' \sigma'|H_3|m\sigma\rangle = 
\gamma \, \langle m'\sigma'| 
             (b k_z^2 - k_x^2 - k_y^2)(k_y \sigma_x - k_x \sigma_y)
          |m\sigma\rangle
\end{equation} 
can be cast in the form
\begin{equation} \label{eq:res_matrix_element_3}
\langle m'\sigma'|H_3|m\sigma\rangle = 
-\frac{\sigma \gamma}{3 a^3} \, 
        \delta_{\sigma',\overline{\sigma}} 
         \, \frac{R_{m'm}}{a} \,
        \sin\theta \ e^{i\sigma \phi} \ e^{-R_{m'm}/a} \ 
        \left\{ 5-b + \left[(b+1) \cos^2\theta -1 \right] \frac{R_{m'm}}{a}
        \right\} \, .
\end{equation}

According to Eqs.~\eqref{eq:res_matrix_element_1} and \eqref{eq:res_matrix_element_3}, 
in the case of the WZ crystal structure we have
$\mathcal{C}_z(r)=0$, and $\mathcal{C}_j(r)=\mathcal{C}^{(1)}_j(r)+\mathcal{C}^{(3)}_j(r)$ for $j=x,y$, with
\begin{subequations} \label{eq:linear_coefficients}
\begin{eqnarray}
\mathcal{C}^{(1)}_x(\mathbf{r})&=&
\frac{\alpha y}{3a^2}\left(1+\frac{r}{a}\right)e^{-r/a} \, , \\
\mathcal{C}^{(1)}_y(\mathbf{r})&=&
-\frac{\alpha x}{3a^2}\left(1+\frac{r}{a}\right)e^{-r/a} \, ,
\end{eqnarray}
\end{subequations}
and
\begin{subequations} \label{eq:cubic_coefficients}
\begin{eqnarray}
\mathcal{C}^{(3)}_x(\mathbf{r})&=&
-\frac{\gamma y}{3a^4}
\left(5-b+\left[(b+1)\left(\frac{z}{r}\right)^2-1\right]\frac{r}{a}\right)e^{-r/a} \, , \\
\mathcal{C}^{(3)}_y(\mathbf{r})&=&
\frac{\gamma x}{3a^4}
\left(5-b+\left[(b+1)\left(\frac{z}{r}\right)^2-1\right]\frac{r}{a}\right)e^{-r/a} \, .
\end{eqnarray}
\end{subequations}

In what follows we calculate the spin-relaxation rate in the above-defined model, using two different 
theoretical approaches.

\section{Semiclassical approach to the spin lifetime}
\label{sec:semiclassical}

The spatial diffusion of electrons through the network of impurities is 
accompanied by a corresponding dynamics of the electronic spin. The spin-orbit 
coupling implies that the spin is not conserved in a hopping event between two 
impurities. In this Section we present the derivation of the spin-relaxation rate 
based on the concept of spin diffusion in the Bloch sphere that occurs when the 
electron undergoes diffusive motion in real space. 
Identifying the electron spin with a continuous vector of fixed norm in 
three-dimensional space amounts to a semiclassical description of the 
spin diffusion.

\subsection{Diffusion on a sphere}

The evolution of the electron spin can be assimilated to a 
random walk on a sphere with a 
mean squared rotation angle $\chi^2$
in each step of the random walk. 
From an initial spin orientation 
$\mathbf{S}_0$ (the electron spin in units of $\hbar/2$) 
corresponding to a distribution of absolute certainty that the point is at a 
given position on the sphere, the resulting distribution after a random walk of 
$N(t)$ steps after a time $t$ is given by \cite{roberts-ursell-1960}
\begin{equation}\label{eq:diff_sph}
\rho(\vartheta,t) = \sum_{n=0}^{\infty} \frac{2n+1}{4\pi} \
      \exp{\left[-\frac{1}{4}n(n+1)V(t)\right]} \ P_n(\cos{\vartheta}) \, .
\end{equation}
Here, $\vartheta$ is the angular distance from the initial orientation, $P_n$ are 
the Legendre polynomials, and $V(t) = N(t) \chi^2$ is the variance of the 
corresponding plane motion. For steps being associated with hopping events 
occurring with a rate $\tau_\mathrm{c}^{-1}$, one has $N(t) = t/\tau_\mathrm{c}$ 
and then $V(t) = (t/\tau_\mathrm{c}) \chi^2$.

The mean projection of the resulting spin orientation $\mathbf{S}(t)$ on the 
initial orientation is given by
\begin{equation}\label{eq:s_z}
\mathbf{S}(t)\cdot\mathbf{S}_0 = 2\pi \int_{0}^{\pi} {\rm d}\vartheta \sin{\vartheta} 
\cos{\vartheta} \, \rho(\vartheta,t) \, .
\end{equation}
With the expression \eqref{eq:diff_sph} and using 
the orthogonality of the Legendre polynomials, one finds an exponential decay of 
the mean spin projection
\begin{equation}
\mathbf{S}(t)\cdot\mathbf{S}_0 = \exp{\left(-\frac{t}{\tau_s}\right)} 
\end{equation}
with the spin-relaxation rate
\begin{equation}
\frac{1}{\tau_\mathrm{s}} = \frac{1}{2} \, \frac{\; \chi^2}{\tau_\mathrm{c}}
\label{eq:reltimedef}
\end{equation}
that depends on the hopping rate and the mean-square spin-rotation angle.

\subsection{Spin rotation in a hopping event}

We consider an initial state where the electron is localized on the impurity $m$ 
with a spin state $|I\rangle$ (not necessarily a state $|\sigma\rangle$ oriented 
along the $z$-axis). 
The initial spin expectation value writes
\begin{equation}
\mathbf{S}_\mathrm{i}=\langle I| \bm{\sigma} |I\rangle \, .
\end{equation}
The final spin state $|F\rangle$ after a hop to impurity $m'$ is obtained as 
\begin{equation}\label{eq:spin_evolution}
|F\rangle = \mathcal{V}(\mathbf{R}_{m'm})|I\rangle \, ,
\end{equation}
where $\mathcal{V}(\mathbf{R}_{m'm})$ is the hopping matrix given by 
Eq. (\ref{eq:spin_matrix_element}). 
The final spin expectation value is computed as 
\begin{equation}
\mathbf{S}_\mathrm{f}=
\frac{\langle F|\bm{\sigma}|F\rangle}{\langle F|F\rangle} \, ,  
\end{equation}
and the spin rotation angle $\alpha_{m'm}$ related to the hopping event can be extracted from 
the scalar product with the initial spin orientation as 
\begin{equation}
\cos\alpha_{m'm} = 
\mathbf{S}_\mathrm{f}\cdot\mathbf{S}_\mathrm{i} \, .
\end{equation}
Using \eqref{eq:spin_evolution} and the general form of the matrix 
elements \eqref{eq:spin_matrix_element}, we calculate the spin rotation. 
To lowest order in the spin rotation components of  $\mathcal{V}(\mathbf{r})$, 
that we assume to be small as compared to the spin-independent hopping terms, 
$\mathcal{C}_j(\mathbf{r}) \ll \mathcal{V}_0(\mathbf{r})$, and therefore 
having small rotation angles 
with $\cos\alpha \approx 1-\alpha^2/2$, we get the result 
\begin{equation}\label{eq:rotangle_initial}
\alpha_{m'm}^2\approx \frac{4}{\mathcal{V}_0^2(\mathbf{R}_{m'm})}
\left\{\mathcal{C}_x^2(\mathbf{R}_{m'm})+
\mathcal{C}_y^2(\mathbf{R}_{m'm})+
\mathcal{C}_z^2(\mathbf{R}_{m'm})
-\left[\mathcal{C}_x(\mathbf{R}_{m'm}) S_\mathrm{i,x}
       +\mathcal{C}_y(\mathbf{R}_{m'm}) S_\mathrm{i,y}+
       \mathcal{C}_z(\mathbf{R}_{m'm}) S_\mathrm{i,z}
 \right]^2 \right\} ,
\end{equation}
which depends on the initial spin orientation $\mathbf{S}_\mathrm{i}$. 
An average over the initial spin orientation based on a homogeneous distribution 
over the surface of the Bloch sphere yields the mean square rotation angle occurring 
in a hopping event 
\begin{equation}\label{eq:spinrotangleif}
\langle\alpha_{m'm}^2\rangle \approx 
\frac{8}{3\mathcal{V}_0^2(\mathbf{R}_{m'm})}
\left[\mathcal{C}_x^2(\mathbf{R}_{m'm})+\mathcal{C}_y^2(\mathbf{R}_{m'm})
      +\mathcal{C}_z^2(\mathbf{R}_{m'm})
\right] \, .
\end{equation}

\subsection{Impurity averaged rotation angle}

In order to use the result of the diffusion on a sphere \eqref{eq:reltimedef}, 
we need the average square spin rotation angle for hops between any pair of impurities. 
Towards the evaluation of that average, we start with the probability that a hop 
from the initial impurity $m$ ends at a given impurity $m'$
\begin{equation}
P_{m'm}=\frac{\mathcal{V}_0^2(\mathbf{R}_{m'm})}
{\sum_{m''\neq m}\mathcal{V}_0^2(\mathbf{R}_{m''m})} \, .
\end{equation}  
We have neglected the small corrections due to the spin-dependent hopping terms $\mathcal{C}_j(\mathbf{r})$. 
Weighting the mean square spin rotation angle \eqref{eq:spinrotangleif} with this 
probability leads to an impurity and spin-orientation averaged squared spin 
rotation angle 
\begin{equation}\label{eq:rotanglesum}
\overline{\langle\alpha^2\rangle} = 
\sum_{m'\neq m} 
P_{m'm} \ \langle\alpha_{m',m}^2\rangle 
=\frac{8}{3} \ \frac{\sum_{m'\neq m}
\left[\mathcal{C}_x^2(\mathbf{R}_{m'm})+
      \mathcal{C}_y^2(\mathbf{R}_{m',m})
      +\mathcal{C}_z^2(\mathbf{R}_{m',m})
\right]}
{\sum_{m'\neq m}\mathcal{V}_0^2(\mathbf{R}_{m'm})} \, .
\end{equation}
We identify the squared mean rotation angle $\chi^2$ used in the formalism of 
the diffusion on a sphere with the average of \eqref{eq:rotanglesum} over 
impurity configurations. 
This impurity average is obtained through the replacement of the sums over final 
positions by integrals over the spatial vector between initial and final site, 
weighted by the constant density of impurities $n_\mathrm{i}$, yielding
\begin{equation}\label{eq:xiintegrals}
\chi^2=\frac{8}{3}\frac{\int\mathrm{d}\mathbf{r}\, n_\mathrm{i}
\left[\mathcal{C}_x^2(\mathbf{r})+\mathcal{C}_y^2(\mathbf{r})
      +\mathcal{C}_z^2(\mathbf{r})
\right]}
{\int\mathrm{d}\mathbf{r}\, n_\mathrm{i}\mathcal{V}_0^2(\mathbf{r})} \, .
\end{equation}
Defining the inverse hopping rate from the time $\tau_\mathrm{c}$ when the 
occupation of a site is halved, \cite{intronati-PRL2012,intronati_thesis} we have 
\begin{equation}\label{eq:hoppingtimeint}
\int\mathrm{d}\mathbf{r}\, n_\mathrm{i}\mathcal{V}_0^2(\mathbf{r})=
\frac{\hbar^2}{2\tau_\mathrm{c}^2} \, .
\end{equation}
Replacing the denominator of \eqref{eq:xiintegrals} using \eqref{eq:hoppingtimeint} 
and plugging the result in the diffusion expression \eqref{eq:reltimedef}, we get the 
mean spin-relaxation rate
\begin{equation}
\left\langle\frac{1}{\tau_\mathrm{s}}\right\rangle = 
\frac{8n_\mathrm{i}\tau_\mathrm{c}}
{3\hbar^2}\int\mathrm{d}\mathbf{r}
\left[\mathcal{C}_x^2(\mathbf{r})+\mathcal{C}_y^2(\mathbf{r})
      +\mathcal{C}_z^2(\mathbf{r})
\right] \, .
\end{equation}

The spin-independent hopping is dominated by $\mathcal{V}_0(\mathbf{r})$ 
[see definition after Eq.~(\ref{eq:spin_matrix_element})]. 
Those terms are the hopping elements of the 
Matsubara-Toyozawa model. \cite{matsubara-1961} 
Using this form in \eqref{eq:hoppingtimeint}, one gets the hopping time\cite{intronati-PRL2012} 
\begin{equation}\label{eq:hoppingtime}
\tau_\mathrm{c}=\frac{\hbar}{\sqrt{14\pi V_0^2 n_\mathrm{i} a^3} } \, ,
\end{equation}
and the general expression of the mean spin-relaxation rate in terms of the model 
parameters as
\begin{equation}\label{eq:reltimegen}
\left\langle\frac{1}{\tau_\mathrm{s}}\right\rangle = 
\frac{8\sqrt{n_\mathrm{i}a^3}}
{3\sqrt{14\pi}\hbar V_0}\int\frac{\mathrm{d}\mathbf{r}}{a^3}
\left[\mathcal{C}_x^2(\mathbf{r})+\mathcal{C}_y^2(\mathbf{r})
      +\mathcal{C}_z^2(\mathbf{r})
\right] \, ,
\end{equation}
with the general property that the relaxation rate is proportional to the 
square root of the dopant density.

\subsection{Application to materials with wurtzite crystal structure}

The expression \eqref{eq:reltimegen} for the spin-relaxation rate is completely
general, only relying on the form \eqref{eq:spin_matrix_element} of the hopping 
amplitudes.
The dominant spin-dependent terms depend on the crystal structure, and  the 
application of the semiclassical spin-diffusion formalism to zincblende materials 
is presented in Appendix \ref{sec:appendixB}.

For wurtzite materials, the spin parts of the hopping matrix elements have 
significant contributions from both, the linear-in-$k$ and the cubic-in-$k$ 
terms in the spin-orbit coupling. 
Using the expressions of Eqs.~(\ref{eq:linear_coefficients}) and (\ref{eq:cubic_coefficients})
in the general formula for the averaged spin-relaxation rate \eqref{eq:reltimegen} 
and performing the integral over hopping vectors, 
one gets with the hopping time \eqref{eq:hoppingtime} the mean 
spin-relaxation time in wurtzite structures
\begin{equation}\label{eq:spinrelaxwu}
\left\langle\frac{1}{\tau_\mathrm{s}}\right\rangle = 
\frac{8\sqrt{\pi}}{\sqrt{14}\,\hbar V_0} 
\sqrt{ n_\mathrm{i} a^3}
E_{\rm soc}^2 \, ,
\end{equation}
where we introduced the material-dependent energy $E_{\rm soc}$ associated to the spin-orbit
coupling and given by
\begin{equation}
E_{\rm soc} \equiv
\left(
      \frac{\alpha^2}{a^2} 
      - \frac{38-6b}{27} \frac{\alpha\gamma}{a^4}
      +\frac{142-38b+9b^2}{189} \frac{\gamma^2}{a^6}
\right)^{1/2} \, .
\label{eq:E_soc}
\end{equation}
The first term is due to the linear-in-$k$ spin-orbit coupling and the second 
and third terms appear due to the presence of the cubic-in-$k$ contribution.

\begin{figure}
\centerline{\includegraphics[width=0.6\textwidth]{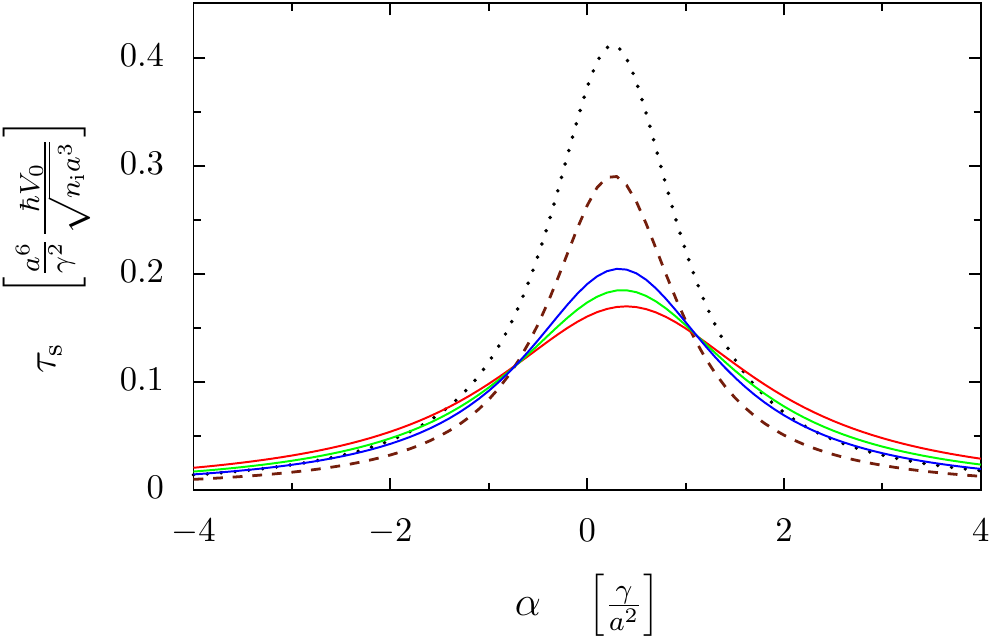}}
\caption{
Spin lifetime as a function of the ratio between the linear-in-$k$ and the cubic-in-$k$ 
spin-orbit coupling parameters, for $b=4$, within the semiclassical approach (black dotted),
the simplest self-consistent approximation (dashed brown),
and the loop-corrected self-consistent approximation (solid).
The first two results are universal in terms of their density dependence 
[see Eqs.~\eqref{eq:spinrelaxwumin} and \eqref{eq:tausssca}], 
while the result for the LCSCA depends on the doping density.
The three solid curves correspond to $n_{\rm i}^{1/3} a = 0.25$ (blue), 
0.29 (green), and 0.33 (red).
\label{fig:lifetime_alpha}}
\end{figure}

As discussed in Sec.\ \ref{sec:hamiltonian}, while  the cubic coupling strength $\gamma$ 
is a material parameter, the linear coupling constant $\alpha$ is composed of 
an intrinsic component $\alpha_{\mathrm{D}}$ 
and an extrinsic Rashba component $\alpha_{\mathrm{R}}$.  
Thus, $\alpha$ can be affected by external influences like strain or an 
electric field, and its value can be controlled and optimized in the search for maximum 
spin lifetime. 
Our general result \eqref{eq:spinrelaxwu} implies a minimal spin-relaxation rate of
 \begin{equation}\label{eq:spinrelaxwumin}
\left\langle\frac{1}{\tau_\mathrm{s}}\right\rangle^\mathrm{min} = 
\frac{8\sqrt{\pi}}{\sqrt{14} \, \hbar V_0}  \frac{\gamma^2}{a^6} 
\sqrt{ n_\mathrm{i} a^3}\frac{1307-228b+180b^2}{5103}
\approx 2.4\frac{\sqrt{ n_\mathrm{i} a^3}}{\hbar V_0}  \frac{\gamma^2}{a^6} 
\end{equation}
occurring for the optimal value $\alpha^\mathrm{opt}$ given by
\begin{equation}\label{eq:alphamin}
\alpha^\mathrm{opt} = \frac{19-3b}{27} \frac{\gamma}{a^2} \approx 0.26 \frac{\gamma}{a^2}  \, ,
\end{equation}
where we have assumed $b=4$ for the evaluation of the approximate numerical 
values.

Defining the spin lifetime as the inverse of the mean relaxation rate  
\eqref{eq:spinrelaxwu}, one gets the Lorentzian dependence on the linear coupling 
strength $\alpha$ depicted in Figure \ref{fig:lifetime_alpha}, with a maximum spin 
lifetime $\tau^\mathrm{max}$ corresponding to the inverse of the 
minimal relaxation rate \eqref{eq:spinrelaxwumin}.  
The relaxation time has 
a pronounced maximum as a function of $\alpha$ with high values close to 
$\alpha^\mathrm{opt}$. 
The width at half maximum of the Lorentzian 
dependence of the lifetime on $\alpha$ is given by
\begin{equation}\label{eq:peakwidth}
\Delta \alpha = 2 \frac{\gamma}{a^2}\sqrt{\frac{1307-228b+180b^2}{5103}} 
\approx 1.6 \frac{\gamma}{a^2} \, . 
\end{equation}

A similar strongly peaked dependence of the spin lifetime on the linear 
coupling has been obtained 
\cite{harmon-APL2011} for conduction-band electrons in wurtzite 
semiconductors based on the Dyakonov-Perel mechanism.   
The presentation of Fig.\ \ref{fig:lifetime_alpha} in terms of dimensionless
quantities is universal and can be used for different materials once their
characteristic constants are determined.

\subsection{Anisotropy of the spin-relaxation rate}

It can however be noticed that the spin-orbit coupling is anisotropic in wurtzite 
structures, with $\mathcal{C}_z(\mathbf{r})=0$. 
If we fix the initial spin orientation along the $z$-axis in \eqref{eq:rotangle_initial}, 
instead of averaging over the initial spin direction, we get the squared mean rotation angle
\begin{equation}
  \chi_{z}^2\approx \frac{4}{\mathcal{V}_0^2(\mathbf{r})}
  \left[\mathcal{C}_x^2(\mathbf{r})+\mathcal{C}_y^2(\mathbf{r})\right] ,
\end{equation}
and thus a (longitudinal) spin-relaxation rate
\begin{equation}
  \frac{1}{\tau_{\mathrm{s},z}}=\frac{3}{2}
  \left\langle\frac{1}{\tau_\mathrm{s}}\right\rangle
\label{eq:spinrelaxwuz}
\end{equation}
that is enhanced by a factor of 3/2. 
In contrast, when the initial spin orientation is in the $x$-$y$ plane, 
a reduced (transverse) spin-relaxation rate of 
\begin{equation}\label{eq:spinrelaxwuxy}
  \frac{1}{\tau_{\mathrm{s},x-y}}=\frac{3}{4}
  \left\langle\frac{1}{\tau_\mathrm{s}}\right\rangle
\end{equation}
is obtained. 
Thus, the initial relaxation is twice as fast for a spin oriented along 
the symmetry axis [001] of the crystal than for a spin perpendicular to that axis.
Such an anisotropy of the spin relaxation does not occur in zinc blende structures 
where the components of the spin-orbit coupling do not have a preferential direction 
(see Appendix \ref{sec:appendixB}).

The factor 2 appearing in the ratio between $\tau_{\mathrm{s},x-y}$ 
and $\tau_{\mathrm{s},z}$ is a general feature of the WZ structure, 
and it is in line with the anisotropic electron spin relaxation measured \cite{bus-etal-2009} 
in bulk GaN at a temperature $T=\unit[80]{K}$, in the regime where the Dyakonov-Perel 
mechanism sets the spin lifetimes. 

\section{Self-consistent approach}
\label{sec:ssa}

The semiclassical approach to the spin lifetime presented in the previous Section 
has been shown to be extremely successful in the case of ZB materials, \cite{intronati-PRL2012} 
as it gives good account of the existing measurements and the results of microscopic theories.
For the low-temperature spin relaxation in the WZ materials, the experimental situation 
is somewhat uncertain, and microscopic theories have not been developed.
The microscopic approach developed for the case of ZB structures \cite{wellens-jalabert-2016} 
uses self-consistent approximations where the locator expansion for the one- and two-particle 
Green functions (and their irreducible components) fulfill important constraints, like particle 
conservation. \cite{vol-woe,rev-vol-woe,kkw90}
We present in this Section different schemes of the self-consistent approximation applicable 
to the WZ crystal structure.
While the general features of the diagrammatic perturbation theory are the same as in the ZB case, 
the reduced symmetry of the WZ structure leads to considerably different results.

The impurity-averaged retarded (advanced) Green function $G^{(\pm)}(\varepsilon)$ can be written 
in terms of the corresponding self-energy $\Sigma^{(\pm)}(\varepsilon)$ through Dyson equation \\
\\
\begin{equation}
G^{(\pm)}(\varepsilon)=\frac{1}{\z_{\pm}-\Sigma^{(\pm)}(\varepsilon)}
\label{eq:sigmadef} \, .
\end{equation} \\
\\
We note $\z_{\pm}=\varepsilon \pm i \eta$, with $\eta$ an infinitesimal positive quantity.
$G^{(\pm)}$ and $\Sigma^{(\pm)}$ are understood as on-site, and the impurity average
makes the choice of the site irrelevant.
Moreover, these matrices are proportional to the $2 \times 2$ identity matrix, and this
is why the spin indices are not explicit.

The intensity propagator $\Phi$ is a two-Green function object that, according to the
Bethe-Salpeter equation, writes as
\begin{equation}
\label{eq:bs00}
\Phi^{\sigma_{1}^{\prime}\sigma_{2}^{\prime},\sigma_{1}\sigma_{2}}
(\varepsilon,\omega,{\bf r}) = 
G^{(+)}\left(\varepsilon_1\right) \ G^{(-)}\left(\varepsilon_2\right) \left[\delta_{\sigma_{1}^{\prime},\sigma_{1}}
\delta_{\sigma_{2}^{\prime}\sigma_{2}}\delta({\bf r})+
\sum_{\sigma_{1}^{\prime\prime}\sigma_{2}^{\prime\prime}}
\int{\rm d}{\bf r}'' \
\Phi^{\sigma_{1}^{\prime}\sigma_{2}^{\prime},
\sigma_{1}^{\prime\prime}\sigma_{2}^{\prime\prime}}
(\varepsilon,\omega,{\bf r}'') \ U^{\sigma_{1}^{\prime\prime}\sigma_{2}^{\prime\prime},\sigma_{1}\sigma_{2}}
(\varepsilon,\omega,{\bf r}-{\bf r}'')\right]
 \, ,
\end{equation} 
in terms of its irreducible component $U$.
We note $\varepsilon=(\varepsilon_1+\varepsilon_2)/2$ and $\hbar \omega=\varepsilon_1-\varepsilon_2$ the semi-sum and the difference of the energies $\varepsilon_1$ and $\varepsilon_2$ of
the two Green functions defining $\Phi$, and $\mathbf{r}$ the vector difference between the
positions of the impurities defining the Green functions.
The interest of the intensity propagator is that it allows us to obtain the probability
distribution governing the spatial and spin diffusion
\begin{equation}
P^{\sigma'\sigma}(\varepsilon,t,{\bf r})=  \frac{n_\mathrm{i}}{\rho(\varepsilon)} \ \frac{\hbar}{2\pi} 
\int_{-\infty}^{+\infty} {\rm d} \omega \ e^{-i\omega t} \
\Phi^{\sigma^{\prime}\sigma^{\prime},\sigma\sigma}(\varepsilon,\omega,{\bf r}) \, ,
\label{eq:defPrtssp}
\end{equation}
where $\rho(\varepsilon)$ denotes the impurity-averaged density of states, obtained as 
\begin{equation}
\label{eq:dos}
\rho(\varepsilon) = 
-\frac{n_\mathrm{i}}{\pi} \ {\rm Im}\left\{G^{(+)}(\varepsilon)\right\} \, .
\end{equation}

From Eqs.~\eqref{eq:linear_coefficients} and \eqref{eq:cubic_coefficients} (together with $C_z=0$) it follows that the hopping amplitude matrix $\mathcal{V}(\mathbf{r})$ defined 
in Eq.~\eqref{eq:spin_matrix_element} fulfills
${\mathcal V}(x,y,z)=e^{i\frac{\phi}{2}\sigma_z} {\mathcal V}(x',y',z) e^{-i\frac{\phi}{2}\sigma_z}$ with 
$x'=\cos(\phi)x-\sin(\phi)y$ and $y'=\sin(\phi)x+\cos(\phi)y$ 
and ${\mathcal V}(x,y,z)=\sigma_y {\mathcal V}(x,-y,z)\sigma_y$.
These transformation properties
dictate that the Fourier transform of the matrix $U$ 
has the following form for ${\bf q}=0$:
\begin{eqnarray}
   \tilde{U}(\varepsilon,\omega,0) =  
   \left(\begin{array}{cccc} \tilde{u}_1(\varepsilon,\omega) & 0 & 0 & \tilde{u}_2(\varepsilon,\omega)\\ 
   0 & \tilde{u}_3(\varepsilon,\omega) & 0 & 0\\ 0 & 0 & \tilde{u}_3(\varepsilon,\omega) & 0 \\ 
   \tilde{u}_2(\varepsilon,\omega) & 0 & 0 & \tilde{u}_1(\varepsilon,\omega)\end{array}\right) \, .
\label{eq:Atilde}
\end{eqnarray}

Restricting ourselves to the two-dimensional subspace of diagonal spin density operators 
(entries $++$ and $--$ of the matrix $\tilde{U}$), we obtain the probability-conserving condition, 
as well as the longitudinal relaxation rate (in the $z$ direction) \cite{wellens-jalabert-2016}
\begin{equation}
   \frac{1}{\tau_{\mathrm{s},z}(\varepsilon)}=\frac{4\pi\rho(\varepsilon)}{ \hbar n_\mathrm{i}} 
   \ \tilde{u}_2(\varepsilon,0) \, .
\label{eq:taus}
\end{equation}
The remaining subspace (entries $+-$ and $-+$) provides the damping of coherences in the chosen representation, 
corresponding to the transverse relaxation rate (in the $x$ and $y$ directions)
\begin{equation}
\frac{1}{\tau_{\mathrm{s},x-y}(\varepsilon)}=\frac{2\pi\rho(\varepsilon)}{ \hbar n_\mathrm{i}} \ 
                                             \Bigl(\tilde{u}_1(\varepsilon,0)+\tilde{u}_2(\varepsilon,0)-\tilde{u}_3(\varepsilon,0)\Bigr) \, .
\label{eq:taustrans}
\end{equation}

The access to the energy-dependent spin-relaxation rates provided by the self-consistent approximation 
is important in order to be able to address not only the case of uncompensated semiconductors 
(with a half-filled impurity band), but also that of weak compensation. 
In addition, it is a necessary information for treating the hot-electron condition that might arise 
from the initial spin injection. \cite{wu-jia-wen}

\subsection{Simplest self-consistent approximation}
\label{subsec:ssca}

In the simplest self-consistent approximation (SSCA) the locator expansion for the self-energy 
is restricted to the only term that represents a processes where the electron hops from site $m$ 
to another site $m''\neq m$, and then back to $m$. Therefore,

\begin{equation}
\Sigma^{(\pm)}(\varepsilon)  =   
\beta \ G^{(\pm)}(\varepsilon) \, , 
\label{eq:sigma00b}
\end{equation}
with
\begin{equation}
\label{eq:beta}
\beta= 
n_{\rm i}\int{\rm d}{\bf r} \mathcal{V}({\bf -r})\mathcal{V}({\bf r})= 
7 \pi \ n_{\rm i}a^3 V_0^2
\left(
      1 + \frac{3 E_{\rm soc}^2 }{7 V_0^2}
\right) \, ,
\end{equation}
where $E_{\rm soc}$ has been defined in Eq.\ \eqref{eq:E_soc}.

The self-consistent retarded self-energy $\Sigma^{(+)}(\varepsilon)$ 
with negative imaginary part reads
\begin{equation}
\Sigma^{(+)}(\varepsilon)=
\frac{1}{2} \left(\z_+- i \sqrt{4\beta-\z_+^2}\right) \, ,
\label{eq:sigma0sol}
\end{equation}
and, according to Eq.~\eqref{eq:dos}, the resulting density of states is given by the semicircle law
\begin{equation}
\rho(\varepsilon) = n_\mathrm{i} \frac{\sqrt{4\beta-\varepsilon^2}}
{2\pi\beta} \ \Theta(4\beta-\varepsilon^2) \, ,
\label{eq:rho0}
\end{equation}
where $\Theta$ stands for the Heaviside door function. 
Since, within this approximation, the impurity band is symmetric around the energy origin, 
the Fermi energy is $\varepsilon_{\mathrm F}=0$

The irreducible component $U$ of the intensity propagator compatible with the simplest 
approximation \eqref{eq:sigma00b} for the self-energy is such that 
\begin{equation}
U^{\sigma^{\prime}_1\sigma^{\prime}_2,\sigma_1\sigma_2}(\varepsilon,\omega,{\bf r})   =   
 n_\mathrm{i}  {\mathcal V}^{\sigma^{\prime}_1\sigma_1}({\bf r})
\left({\mathcal V}^{\sigma^{\prime}_2\sigma_2}({\bf r})\right)^* \, ,
\label{eq:A00}
\end{equation}
and it can thus be expressed as the tensor product $U(\varepsilon,\omega,{\bf r})  =   
n_\mathrm{i}{\mathcal V}({\bf r})\otimes{\mathcal V}^*({\bf r})$. 
In Fourier space we have
\begin{equation}
\tilde{U}(\varepsilon,\omega,{\bf q})  = n_\mathrm{i}  \int\frac{{\rm d}{\bf k}} {(2\pi)^3} \ 
\tilde{\mathcal V}({\bf k}_+)\otimes \tilde{\mathcal V}^*({\bf k}_-) \, ,
\label{eq:A0}
\end{equation}
where $\tilde{\mathcal V}({\bf k})$ stands for the Fourier transform of the hopping amplitude 
matrix given in Appendix \ref{app:ftham}, and ${\bf k}_\pm={\bf k}\pm{\bf q}/2$.

According to Eqs.~\eqref{eq:Atilde} and \eqref{eq:A00},
\begin{subequations}
\begin{eqnarray}
\tilde{u}_1 & = & \tilde{u}_3 = 
n_{\rm i} \int{\rm d}{\bf r} \ \mathcal{V}_0^2({\bf r}) = 
7 \pi n_{\rm i} a^3  V_0^2 \, , \\
\tilde{u}_2 & = & n_{\rm i}\int{\rm d}{\bf r} \ 
\mathcal{V}^{\overline{\sigma},\sigma}({\bf r})
\left(\mathcal{V}^{\sigma,\overline{\sigma}}({\bf r})\right)^* = 
3 \pi n_{\rm i} a^3 E_{\rm soc}^2 \, .
\end{eqnarray}
\end{subequations}

From \eqref{eq:taus}, \eqref{eq:beta}, and \eqref{eq:rho0} the longitudinal relaxation rate 
for electrons at the Fermi energy $\varepsilon_{\mathrm F}$ is 
\begin{equation}
  \frac{1}{\tau_{\mathrm{s},z}(\varepsilon_{\mathrm F})}=
  \frac{12\sqrt{\pi}}{\sqrt{7}\hbar V_0} \sqrt{ n_\mathrm{i} a^3} \
  \frac{E_{\rm soc}^2}{\left[1 + \frac{3 E_{\rm soc}^2 }{7V_0^2}\right]^{1/2}} \, ,
\label{eq:tausssca}
\end{equation}
and the transverse relaxation rate is
\begin{equation}
\frac{1}{\tau_{\mathrm{s},x-y}(\varepsilon_{\mathrm F})} = \frac{1}{2} \
\frac{1}{\tau_{\mathrm{s},z}(\varepsilon_{\mathrm F})}
\label{eq:transv-long}
\end{equation}
The term that multiplies $1/V_0^2$ in the denominator of Eq.~\eqref{eq:tausssca} stems 
from the small correction of the density of states due to the spin-orbit coupling. 
Neglecting such a term, and taking into account the numerical factors of 
Eqs.~\eqref{eq:spinrelaxwuz}-\eqref{eq:spinrelaxwuxy} relating the averaged spin-relaxation 
rate with the longitudinal and transverse ones, we see that the semiclassical and 
the simplest self-consistent approaches yield the same functional form for the relaxation rates, 
with only a difference of $1/\sqrt{2}$ between the prefactors. 
Such a difference is not surprising, since in the semiclassical approach the exact definition
of the relaxation time is to some extent arbitrary. 
For instance, the hopping time $\tau_\mathrm{c}$ is defined in \eqref{eq:hoppingtimeint} 
as the time when the occupation of a site is halved while a somewhat different criterion 
would have been equally adequate.
Within the diagrammatic approach, the hopping time can be expressed in terms of the imaginary part 
of the self energy as follows:\cite{wellens-jalabert-2016}
\begin{equation}
\tau_\mathrm{c}=-\frac{\hbar}{2{\rm Im}\left\{\Sigma^{(+)}(\varepsilon)\right\}}
\label{eq:taucdiagrammatic}
\end{equation}
which, when taking into account Eqs.~\eqref{eq:beta} 
and \eqref{eq:sigma0sol}, 
differs from \eqref{eq:hoppingtimeint} 
again by a factor $\sqrt{2}$. 
In total, the simplest self-consistent approximation exactly reproduces the semiclassical approach 
with hopping time $\tau_\mathrm{c}$ defined according to \eqref{eq:taucdiagrammatic}.

\subsection{Loop-corrected self-consistent approximation}
\label{subsec:lcsca}

The simplest self-consistent approximation developed in Subsec.\ \ref{subsec:ssca} needs to be 
complemented by adding the terms of the locator expansion that represent processes 
in which the electron visits more than one impurity before hopping back to the starting one. 
Such an improvement constitutes the so-called loop-corrected self-consistent approximation (LCSCA), 
\cite{wellens-jalabert-2016} which for the spinless case, amounts to the approach used by
Matsubara and Toyozawa \cite{matsubara-1961} in order to obtain the density of states 
in the impurity band, as well as the conductivity within the diffusion approximation. 

Since the self-energy is now expressed as a geometrical series representing hopping events, 
Eq.~\eqref{eq:sigmadef} can be written as a self-consistent equation for $\Sigma^{(\pm)}(\varepsilon)$
\begin{equation}
\Sigma^{(\pm)}(\varepsilon) =  n_\mathrm{i} \int\frac{{\rm d}{\bf k}}{(2\pi)^3} \ 
\frac{\left(\tilde{\mathcal V}^{(\pm)}({\bf k})\right)^2}{\z_{\pm}-\Sigma^{(\pm)}(\varepsilon)- n_\mathrm{i} 
\tilde{\mathcal V}^{(\pm)}({\bf k})} \, ,
\label{eq:Sigmadscb}
\end{equation}
where we note 
$\tilde{\mathcal V}^{(+)}({\bf k})=\tilde{\mathcal V}({\bf k})$,
$\tilde{\mathcal V}^{(-)}({\bf k})=\tilde{\mathcal V}^{*}({\bf k})$.
This equation needs to be numerically solved. 
Since the spin-orbit coupling only gives a small correction to the density of states, 
we solve for $\Sigma^{(\pm)}(\varepsilon)$ in Eq.~\eqref{eq:Sigmadscb} trading 
$\tilde{\mathcal V}^{(\pm)}({\bf k})$ by $\tilde{\mathcal V}_0({\bf k})$.

The loop-corrected sequence translates into a renormalized hopping amplitude matrix
\begin{equation} \label{eq:F}
\tilde{\mathcal F}^{(\pm)}(\varepsilon,{\bf k}) = 
\frac{\tilde{\mathcal V}^{(\pm)}({\bf k})}{\mathbb{I}_{2} - n_\mathrm{i} \ G^{(\pm)}(\varepsilon) \ 
\tilde{\mathcal V}^{(\pm)}({\bf k})} \, ,
\end{equation}
with $\mathbb{I}_{2}$ the $2\times 2$ unit matrix.
Thus, the irreducible component of the intensity propagator takes the form \eqref{eq:A0}, 
but with the hopping amplitude matrix now replaced by the effective one. \cite{wellens-jalabert-2016}
That is,
\begin{equation}
\tilde{U}(\varepsilon,\omega,{\bf q})  =   n_\mathrm{i}  \int\frac{{\rm d}{\bf k}}{(2\pi)^3} \ 
\tilde{\mathcal F}^{(+)}(\varepsilon_1,{\bf k}_+)\otimes 
\tilde{\mathcal F}^{(-)}(\varepsilon_2,{\bf k}_-) \, .
\label{eq:Fmatrix}
\end{equation}
Neglecting the spin-orbit flipping terms in the denominator of \eqref{eq:F}  results in
\begin{equation}
\tilde{u}_2(\varepsilon,0)   =     
n_\mathrm{i} \int\frac{{\rm d}{\bf k}}{(2\pi)^3} \ 
\frac{|\tilde{\mathcal C}_x({\bf k})|^2+|\tilde{\mathcal C}_y({\bf k})|^2}
{\left|1- n_\mathrm{i} G^{(+)}(\varepsilon)\tilde{\mathcal V}_0({\bf k})\right|^4} \, .
\end{equation}
Performing the angular integrals 
\begin{eqnarray}
\tilde{u}_2(\varepsilon,0)  & = &    
\frac{1024}{945} \ 
n_{\rm i} \gamma^2 a^2 \int_0^\infty {\rm d}k \ 
\frac{1}
{\left|1- n_\mathrm{i} G^{(+)}(\varepsilon)\tilde{\mathcal V}_0(k) \right|^4} \
\frac{k^4}{\left(1 + k^2 a^2 \right)^8} 
\nonumber \\
& & \times \left\{35 \left(6 \frac{\alpha a^2}{\gamma} - 1 \right)^2 
                 -14 k^2 a^2 (29-6 b) \left(6 \frac{\alpha a^2}{\gamma} - 1 \right) 
                 + k^4 a^4 \Bigl[1235-12 (31-9 b)b \Bigr] 
           \right\} \, ,
\label{eq:u2tilde}
\end{eqnarray}
and the longitudinal spin-relaxation rate follows from Eq.~\eqref{eq:taus} 
after the $k$-integration.
From \eqref{eq:u2tilde} we readily see that, similarly to the SSCA, the spin-relaxation
time within the LCSCA has a Lorentzian dependence on the variable  $\alpha a^2/\gamma$.
However, contrary to the simpler approximation, the density dependence is not
universal in the parameter $n_i a^3$.
As in the SSCA approximation, we have $\tilde{u}_1 = \tilde{u}_3$, 
and therefore the same relationship \eqref{eq:transv-long} between the longitudinal 
and the transversal relaxation times.

In Fig.~\ref{fig:lifetime_alpha} we present the LCSCA spin-relaxation times 
in dimensionless variables for three different impurity densities (solid lines).
For low values of $\alpha a^2/\gamma$, the relaxation time decreases as the doping density
increases.
Such a behavior is in line with that of the ZB case in the metallic side of 
the MIT. \cite{intronati-PRL2012,wellens-jalabert-2016}
The optimal values of  $\alpha a^2/\gamma$ that maximize the spin-relaxation time
are close to those of the simpler approximations.
Interestingly, the linear-in-$k$ term of the WZ structure induces a cross-over value
beyond which the relaxation times increase with impurity density.

\subsection{Repeated-scattering-corrected self-consistent approximation}

The LCSCA can be improved by the inclusion of cross diagrams 
in the locator expansion for the self-energy 
that describe the repeated scattering from selected impurities. 
The so-called repeated-scattering-corrected self-consistent 
approximation (RSCSCA) \cite{wellens-jalabert-2016} 
restricts the repeated scattering to just a pair of impurities, 
allowing for arbitrary loops between them 
[represented by the renormalized hopping amplitude \eqref{eq:F}]. 
The irreducible component of the intensity propagator in the RSCSCA has an expression 
considerably more complicated than that of Eq.~\eqref{eq:Fmatrix} since, 
even restricting to a pair of repeatedly visited impurities, there is an important 
proliferation of contributing diagrams.

The RSCSCA results (not shown) are very close to those of the LCSCA, leaving aside 
the high-energy tail of the impurity band, as well as the very low densities, 
for which the repeated scattering is more relevant. \cite{Elyutin}
Since our approach is restricted to impurity densities larger than the critical 
one for the metal-insulator transition, the latter case does not apply to our study. 
Since we work with uncompensated or weakly compensated samples, the first case 
is also not relevant. 

It is interesting to remark that, while the quantitative improvements of the RSSCA respect
to the LCSCA are generically small in the range of parameters that we work with,
the magnitude of these corrections depends on the value of the linear-in-$k$ $\alpha$
coefficient.
For small values of $\alpha$, where the spin-relaxation time is close to $\tau_s^{max}$,
the corrections are of the order of 10\%, while for larger values of $\alpha$
the correction practically vanishes.
The latter result is a consequence of the fact that for large $\alpha$ 
the integral leading to $\tilde{u}_2$ is dominated by the small $k$-values, 
which in turn are associated with large values of $r$, 
where the repeated scattering is not relevant.

\section{Spin-relaxation times in G\lowercase{a}N, Z\lowercase{n}O, I\lowercase{n}N, and A\lowercase{l}N}
\label{sec:qr}

Direct comparison with the few available low-temperature data of spin-relaxation times 
in WZ materials is hindered by the limited knowledge of some material parameters
and the uncertainty on certain experimental conditions, like the excess doping-density 
beyond the critical one or the tuning of the linear-in-$k$ spin-orbit coupling constant 
induced by an electrostatic potential. 
Moreover, the exact nature of the MIT in wide gap WZ doped semiconductors has rarely been   
experimentally investigated, other than in the case of GaN.\cite{wol-etal}  
For these materials, the standard Mott criterion, as well as more refined calculations, 
\cite{fer-per,fer-ara} only provide a qualitative estimate of the critical densities.

Despite these limitations, we verify that our theoretical model 
yields the appropriate order of magnitude of the spin-relaxation times for particularly 
important materials.
We then analyze the improvements that can be made on the spin-relaxation time 
by adjusting the tunable linear-in-$k$ coupling constant $\alpha$, along the lines of
the proposal made in Ref.\ [\onlinecite{wang-JAP2010}] to reduce Dyakonov-Perel relaxation
in wurtzite quantum wells.
In Table \ref{tab:values} we summarize the used material parameters and present results 
for the spin-relaxation times obtained within the framework of the LCSCA. 
The differences with the other calculational schemes are not particularly important, 
especially in the regime of large $\alpha a^2/\gamma$, where the linear-in-$k$ term dominates 
the spin relaxation.

\begin{table}
\caption{\label{tab:values} 
   Material parameters and resulting spin-relaxation times. 
   The effective masses $m^*$ and the dielectric constants $\epsilon$ are taken 
   from Ref.\ \onlinecite{Hanada2009}, while the spin-orbit interaction coefficients 
   $\alpha$, $\gamma$, and $b$ are extracted from the quoted references. 
   The spin lifetime $\tau_s$ and the maximum spin lifetime $\tau_s^\mathrm{max}$ 
   are calculated within the loop-corrected self-consistent approximation (LCSCA) 
   of Sec.~\ref{subsec:lcsca} at the density that corresponds to the Mott criterion 
   $n_\mathrm{i}^{1/3}a\approx 0.25$ for the metal-insulator transition. 
   For the former, the tabulated value of $\alpha$ is used, while for the latter 
   the optimal value $\alpha^{\mathrm{opt}}$ arising from the LCSCA is adopted.}
\begin{ruledtabular}
\begin{tabular}{ccccccccccc}
  Material & $m^*$ & $\epsilon$ & $a$ [$\AA$] & $V_0$ [meV] & $\alpha$ [meV$\AA$] & $\gamma$ [meV$\AA^3$]
           & $b$ & $\tau_s$ [ns] & $\alpha^{\mathrm{opt}}$ [meV$\AA$] &  $\tau_s^\mathrm{max}$ [ns] \\
GaN & 0.19 & 10.1 & 29 & 50  & 4.5\cite{ste-2014}       & 400\cite{yin-APL2010}    & 3.959\cite{fu-wu}        & 5   & 0.2  & 150 \\
ZnO & 0.22 & 7.8  & 19 & 98  & 2.2\cite{and-jar-gra}    & 320\cite{fu-wu}          & 3.855\cite{fu-wu}        & 15  & 0.35 & 38 \\ 
InN & 0.10 & 15.0 & 77 & 13  & 13.1\cite{wang-JAP2010}  & 354\cite{wang-JAP2010}   & 4.885\cite{wang-JAP2010} & 1   & 0.02 & $1.8\times 10^4$ \\  
AlN & 0.28 & 8.5  & 16 & 104 & -0.72\cite{wang-JAP2010} & 6.445\cite{wang-JAP2010} & 3.767\cite{wang-JAP2010} & 118 & 0.01 & $4.0\times 10^4$ \\ 
\end{tabular}
\end{ruledtabular}
\end{table}

The quoted effective masses $m^*$ and dielectric constants $\epsilon$ are direction-averaged values, 
{\it i.e.} $m^*=(m^{\parallel}+2 m^{\perp})/3$, with $m^{\parallel}$ and $m^{\perp}$ being, 
respectively, the longitudinal and transverse effective masses 
with respect to the c-axis of the WZ structure. \cite{Hanada2009} 
These parameters determine the effective Bohr radius $a$ and the spin-independent 
hopping amplitude $V_0$, according to $a=a_0\epsilon/m^*$ and $V_0=2E_\mathrm{R}^{(0)}m^*/\epsilon^2$, 
where $a_0$ and $E_\mathrm{R}^{(0)}$ are, respectively, the bare values of the Bohr radius and 
the Rydberg energy of an isotropic hydrogen atom. \cite{Rodina2001}

The values of the spin-orbit interaction coefficients $\alpha$, $\gamma$, and $b$, were taken 
from the quoted references. 
Whenever known, the experimentally determined values were used. 
For instance, in the case of GaN we chose for the cubic-in-$k$ spin-orbit coupling constant 
the experimental value \cite{yin-APL2010} $\gamma=\unit[400]{meV \AA^3}$, 
instead of the prediction arising from tight-binding band-structure calculations \cite{fu-wu} 
$\gamma=\unit[330]{meV \AA^3}$. 
Similarly, for the linear-in-$k$ spin-orbit coupling constant we use the value $\alpha=\unit[4.5]{meV \AA}$ 
determined from weak antilocalization measurements, \cite{ste-2014} 
rather than the result $\alpha=\unit[9]{meV \AA}$ arising from {\it ab initio} 
computations. \cite{maj-2005-APP, maj-2005} 
In the case of ZnO we used the experimental value \cite{and-jar-gra}  $\alpha = \unit[2.2]{meV \AA}$
instead of $\alpha = \unit[1.1]{meV \AA}$ given by band-structure calculations. \cite{lew-wil-car-chr}

The values of $\tau_s$ in Table \ref{tab:values} are extracted from the data of Fig.~\ref{fig:lifetime_alpha}, 
using the material-dependent physical parameters and setting the doping density to the critical one, 
according to the Mott criterion $n_\mathrm{i}^{1/3}a\approx 0.25$. 
The column $\alpha^{\mathrm{opt}}$ indicates the value of the linear-in-$k$ coupling constant 
for which the relaxation time is maximum. 
%
The last column $\tau_s^\mathrm{max}$ gives the value of the spin-relaxation time 
obtained when the linear-in-$k$ coupling constant takes the value $\alpha^{\mathrm{opt}}$, 
and sets up an upper bound for the times than can be achieved by tuning the contribution arising 
from $\alpha_{\mathrm{R}}$ by the application of an external voltage in the $z$-axis direction.
The electrical tuning of the Rashba spin-orbit interaction can be very large. 
In particular two-fold \cite{sch-zol-fue-mad-nyg-cso} and six-fold \cite{lia-xua} tuning of the Rashba coefficient 
have been reported in InAs nanowires.
In the case of bulk and epilayer wurtzite materials, 
estimating the extent of the tuning span is not simple and remains a task for future investigations.
The values of $\tau_s^{\rm max}$ listed in Table~\ref{tab:values} therefore indicate potential improvements, 
realizable under the condition that $\alpha$ can be tuned to the required value $\alpha^{\mathrm{opt}}$.
The width $\Delta\alpha$ at half maximum of the Lorentzian dependence of the lifetime, 
given approximately by Eq.~\eqref{eq:peakwidth}, indicates how close $\alpha$ has to be 
to $\alpha^{\mathrm{opt}}$ in order to obtain a lifetime that is of the order of 
the optimal value $\tau_s^{\rm max}$.

For GaN, the calculated $\tau_s$ is of the same order as the measured maximum 
spin-relaxation time of around 20~ns at low temperature, \cite{wol-etal,bes-etal} 
and the discrepancy between them could be due to the uncertainty in the precise value 
of $\alpha$. \cite{ste-2014,maj-2005-APP,maj-2005,wol-2011}
In particular, the large number of dislocations reported by Ref.~\onlinecite{bes-etal} 
might affect the anisotropy of the crystal and thereby the linear spin-orbit coupling. 
Furthermore, we see that tuning $\alpha$ from $4.5$~meV$\AA$ to $\alpha^{\rm opt}=0.2$~meV$\AA$ 
leads to a 30-fold increase of the spin relaxation time related to the mechanism under study.

ZnO is the other material for which low-temperature spin-relaxation experimental 
data is available. 
Values of $\tau_s$ around $20~\mbox{ns}$ at $T=30~\text{K}$ have been reported, \cite{gho-etal}
albeit for a sample with a density of $1.26\times 10^{15} \text{cm}^{-3}$, which is very low 
compared to the critical MIT density. 
With this caveat, we remark the good agreement between our theoretical
prediction and the measured value of the spin lifetime in ZnO.
The comparatively larger theoretical estimation of $\tau_s$ in ZnO with respect to GaN is essentially due to the 
fact that, in the case of ZnO, the value of $\alpha$ is closer to the optimum value $\alpha^{\rm opt}$ 
already without external field and, in addition, the width $\Delta\alpha$ at half maximum of 
the Lorentzian dependence of the lifetime, see Eq.~\eqref{eq:peakwidth}, 
is larger ($\Delta\alpha=1.4$ meV$\AA$ for ZnO as compared to $\Delta \alpha = 0.8$ meV$\AA$ for GaN). 
On the other hand, the upper bound $\tau_s^{\rm max}$ for ZnO is considerably smaller than for GaN, 
mainly due to the proportionality of $\tau_s^{\rm max}$ to $a^6$ 
(see the universal scaling indicated in the unit of $\tau_s$ in Fig.~\ref{fig:lifetime_alpha}).

InN and AlN represent extreme cases in terms of their spin-orbit parameters.
While the value of $\alpha_\mathrm{D}$ of InN is large, the one corresponding to AlN 
is very small, and close to $\alpha^{\mathrm{opt}}$.
Furthermore, AlN is peculiar in the sense that the theoretical value \cite{wang-JAP2010} 
for the cubic coupling $\gamma=\unit[6.445]{meV \AA^3}$ is very small in comparison with 
the values of other materials discussed.
Note that, for both materials, the optimized spin-relaxation time $\tau_s^\mathrm{max}$ 
arising from the spin-orbit mechanisms described in Eqs.~\eqref{eq:Hallunrestr}-\eqref{eq:HDress} 
is extremely high.
In the case of InN this is mainly due to the large value of the Bohr radius $a$, while for AlN 
it is due to the small magnitude of $\gamma$.
Thus, these two materials are promising candidates to obtain long 
spin-relaxation times, especially AlN which would require only a weak tuning of $\alpha$. 
The small values of the optimization window $\Delta \alpha$ for InN and AlN, 
$0.1$ and $0.04$ meV$\AA$ respectively, translate into very narrow and tall peaks 
for the $\alpha$-dependence of the spin-relaxation time. 
A similar conclusion was obtained from a theoretical study of spin relaxation in the conduction band 
at high temperatures, \cite{harmon-APL2011} where a very high and narrow peak was found in 
the $\alpha$-dependence of the spin lifetime, with a maximum value of up to \unit[0.5]{$\mu$s} 
at room temperature.
To the best of our knowledge, no measurements of the low-temperature spin-relaxation time 
near the metal-insulator transition in the impurity band have been reported for these 
two interesting semiconductors.

\section{Conclusions}
\label{sec:conclusion}

We have theoretically studied spin relaxation in the metallic regime of the impurity band
in semiconductors with wurtzite crystal structure.
We adapted theoretical concepts previously developed and successfully applied in the context 
of zinc blende semiconductors.
Our basic model is solved using two approaches, namely, a semiclassical one
and a fully quantum-mechanical microscopic theory.
The latter is pursued at three levels of self-consistent approximation: 
the simplest (SSCA), the loop-corrected (LCSCA), and the repeated-scattering-corrected (RSCSCA),
incorporating progressively higher-order terms of the spin-orbit interaction
in a diagrammatic locator expansion.

The anisotropic nature of the wurtzite structure gives rise to a corresponding 
anisotropy in the spin-relaxation time.
The interplay between the linear- and cubic-in-$k$ terms of the spin-orbit Hamiltonian 
leads to a richer scenario of physical behavior, as compared with the case of 
zinc blende materials which only have cubic terms.
The obtained theoretical results are generically expressed in terms of
material-dependent parameters and can thus be applied to different
wurtzite semiconductors of practical interest.
We emphasize that our theory contains no adjustable parameters, and all 
comparisons with experiment have been done using the most reliable material
parameters available in the literature.

The theoretical estimates of the spin-relaxation time obtained with the LCSCA 
were shown to be of the same order of magnitude as the available experimental 
data on GaN and ZnO.
At the quantitative level, while the theory underestimates somewhat the 
spin lifetime for GaN found in experiments, it gives a fairly accurate 
result for ZnO. 
The discrepancy between theory and experiment for GaN calls for further investigation 
of this important material.
Although the scenario for ZnO looks at this point more consistent than that for GaN, 
experimental data for ZnO with densities closer to the MIT would be needed 
to draw definite conclusions.
With an eye to potential spintronic applications, we also analyzed the maximization 
of the spin-relaxation time made possible by adjusting the linear-in-$k$ term of 
the spin-orbit interaction, and showed that radical improvements could be made 
for GaN and AlN, leading to lifetimes given by the discussed mechanism that are 
so long that another mechanism can be expected to dominate the spin relaxation.

\acknowledgments
We gratefully acknowledge support from 
the French National Research Agency ANR (Projects ANR-11-LABX-0058\_NIE and ANR-14-CE36-0007-01), 
the French-Argentinian collaborative project PICS 06687,
Universidad de Buenos Aires (Project UBACyT 2011-2014 No.\ 20020100100741), 
and CONICET (Project PIP 11220110100091).

\appendix

\section{Calculation of hopping matrix elements}
\label{sec:appendixA}

In this Appendix we calculate analytically the matrix elements 
of the spin-orbit Hamiltonian between hydrogenic impurity states.
For the linear-in-$k$ term we change the integration variables in Eq.~\eqref{eq:H1b} 
by shifting the origin of coordinates to $\mathbf{R}_m$,
and we call $\mathbf{R}_{m'm} = \mathbf{R}_{m'} - \mathbf{R}_m$, obtaining
\begin{eqnarray} 
\langle m'\sigma'|H_1|m\sigma\rangle &=& \frac{\sigma \alpha}{a} \, 
                                      \delta_{\sigma' \overline{\sigma}}
   \int \dint \mathbf{r} \, 
             \phi(\mathbf{r}-\mathbf{R}_{m'm}) \ \phi(r) \
             \left(  
             \frac{x+i\sigma  y}{r}
             \right) \nonumber \\
             &=& 
   \frac{\sigma \alpha}{a} \, \delta_{\sigma' \overline{\sigma}} \,
   \left(I_{x,m'm} + i\sigma  I_{y,m'm}\right) \, .
\label{eq:matrix_element_1}
\end{eqnarray}
We have defined
\begin{subequations} 
\begin{eqnarray} I_{x,m'm} &=& \int \dint \mathbf{r} \, 
             \phi(\mathbf{r}-\mathbf{R}_{m'm}) \ \phi(r) \,  
             \frac{x}{r} =
           \frac{1}{\pi a^3} 
           \int \dint \mathbf{r} \, 
             e^{-|\mathbf{r}-\mathbf{R}_{m'm}|/a} \, e^{-r/a} \  
             \frac{x}{r} 
\, ,
\label{subeq:Ix} 
\\
I_{y,m'm} &=& \int \dint \mathbf{r} \, 
             \phi(\mathbf{r}-\mathbf{R}_{m'm}) \ \phi(r) \,  
             \frac{y}{r} =
           \frac{1}{\pi a^3}
            \int \dint \mathbf{r} \, 
             e^{-|\mathbf{r}-\mathbf{R}_{m'm}|/a} \, e^{-r/a} \  
             \frac{y}{r} \, .
\label{subeq:Iy} 
\end{eqnarray}            
\end{subequations}
We write $\mathbf{R}_{m'm} = (X_{m'm},Y_{m'm},Z_{m'm}) =
R_{m'm} (\sin\theta \cos\phi \, \mathbf{x} + \sin\theta \sin\phi \, 
\mathbf{y} + \cos\theta \, \mathbf{z})$, and then Eq.~\eqref{subeq:Ix} takes the form
\begin{equation}
\label{eq:Ix2}
I_{x,m'm} = \frac{1}{\pi a^3} 
           \int \dint \mathbf{r} \, 
             e^{-\sqrt{ (x-X_{m'm})^2 + (y-Y_{m'm})^2 + (z-Z_{m'm})^2}/a } \ 
             e^{-\sqrt{ x^2 + y^2 + z^2}/a } \,  
             \frac{x}{\sqrt{ x^2 + y^2 + z^2 } } \, ,
\end{equation}
The rotation of the coordinate system by $(\theta,\phi)$ induces a coordinate change 
characterized by the transformation
\begin{equation}
 \left( {\begin{array}{c} 
          x \\ y \\ z \\ 
         \end{array} } 
 \right) =
 \left( {\begin{array}{ccc}
          \cos\theta \cos\phi & - \sin\phi & \sin\theta \cos\phi  \\
          \cos\theta \sin\phi  &  \cos\phi & \sin\theta \sin\phi  \\
          -\sin\theta & 0 &  \cos\theta \\
         \end{array} } 
 \right)
 \left( {\begin{array}{c} 
           x' \\ y' \\ z' \\ 
         \end{array} } 
 \right) \, .
\label{eq:inverse_rotation}
\end{equation}
Performing the corresponding change of coordinates, the integral in \eqref{eq:Ix2} becomes
\begin{equation}
I_{x,m'm} = \frac{1}{\pi a^3} 
           \int \dint \mathbf{r}' \, 
             e^{-\sqrt{ x'^2 + y'^2 + (z'-R_{m'm})^2}/a } \ 
             e^{-r'/a} \,  
             \left(\frac{x' \cos\theta \cos\phi- y' \sin\phi + z' \sin\theta \cos\phi}{r'}\right) .
\end{equation}
The cylindrical symmetry around the $z'$-axis calls for a further change of variables in favor 
of the cylindrical coordinates $(\rho,\alpha,z)$:
\begin{equation}
I_{x,m'm} = \frac{1}{\pi a^3} 
           \int \dint z \ \dint \rho \ \dint \alpha \ \rho \ 
             e^{-\sqrt{\rho^2+(z-R_{m'm})^2}/a} \ 
             e^{-\sqrt{\rho^2+z^2}/a} \left(  
             \frac{\rho \cos\alpha \cos\theta \cos\phi- \rho \sin\alpha \sin\phi 
              + z \sin\theta \cos\phi}{\sqrt{\rho^2+z^2}}\right) \, .
\end{equation}
The integration over $\alpha$ only leaves the last term, and then
\begin{equation}
I_{x,m'm} = \frac{2}{a^3}  \sin\theta \ \cos\phi
           \int \dint z \ \dint \rho  \ 
              e^{-\sqrt{\rho^2+(z-R_{m'm})^2}/a} \ 
              e^{-\sqrt{\rho^2+z^2}/a} \,  
             \frac{\rho z}{\sqrt{\rho^2+z^2}} = 
             2  \sin\theta \ \cos\phi \ I_1(R_{nm}/a) .
\end{equation}
Analogously we get 
\begin{equation}
   I_{y,m'm} = \frac{2}{a^3}  \sin\theta \ \sin\phi \, I_1(R_{nm}/a) \, ,            
\end{equation}
and collecting both terms in Eq.\ (\ref{eq:matrix_element_1}) we get
\begin{eqnarray} 
\langle m'\sigma'|H_1|m\sigma\rangle &=&  
   \frac{2\sigma \alpha}{a} \, 
   \delta_{\sigma' \overline{\sigma}} \,
    \sin\theta \, (\cos\phi + i \sigma \sin\phi) 
    \, I_1(R_{nm}/a) \, .
\end{eqnarray}
The integral $I_1$ can be solved exactly:
\begin{equation}
   I_1(\xi) = \frac{1}{6} \ (\xi + \xi^2) \ e^{-\xi} \, ,
\end{equation}
leading to the matrix element of the linear Dresselhaus spin-orbit coupling  
of Eq.~\eqref{eq:res_matrix_element_1}.



The cubic-in-$k$ matrix elements \eqref{eq:res_matrix_element_3f} can be written as
\begin{equation}
\langle m' \sigma'|H_3|m\sigma\rangle = \gamma
\delta_{\sigma',\overline{\sigma}} \,
 \langle m'| \,b\, k_y k_z^2 - i \sigma b k_x k_z^2  
            - k_x^2 k_y + i \sigma k_x^3
            - k_y^3     + i \sigma k_x k_y^2 \,
 |m\rangle \, .
\end{equation}
We have six terms of the form $\langle n|k_i k_j^2|m\rangle$, where
$i,j=x,y,z$ ($i$ and $j$ may be equal).
After acting with the differential operators on the hydrogenic states we get
\begin{eqnarray} 
\langle m' |k_i k_j^2|m\rangle &=& 
   \frac{i}{a^2} \  
   \langle m'| \frac{(x_i-X_{m'i})(x_j-X_{mj})^2}
                   {|\mathbf{r}-\mathbf{R}_{m'}|
                    |\mathbf{r}-\mathbf{R}_m|^2}
              \left(  \frac{1}{a}
                    + \frac{1}{|\mathbf{r}-\mathbf{R}_m|} 
                    - \frac{|\mathbf{r}-\mathbf{R}_m|}{(x_j-X_{mj})^2}
              \right)
   |m\rangle \, .
\end{eqnarray}
The remaining calculation is lengthier than the one of the linear-in-$k$ matrix element, 
but it follows the same steps: 
i) a shift of the origin of coordinates to $\mathbf{R}_m$, 
ii) a rotation of the coordinate system, 
iii) a switch to cylindrical coordinates.
Rather than writing all the details of this procedure, we just make explicit the analytical 
expression of some useful two-dimensional integrals:
\begin{subequations}
\begin{eqnarray}
I^{p,q}(\xi) &=& \frac{1}{a^{p+q-1}}
\int \dint z \ \dint \rho \ 
\frac{e^{-\sqrt{\rho^2+ (z-\xi a)^2}/a} \ e^{-\sqrt{\rho^2+ z^2}/a}}
{\sqrt{\rho^2+ (z-\xi a)^2} \ (\rho^2+ z^2)^{3/2}}
                              \left(a+\sqrt{\rho^2+ z^2}\right)
                              \ \rho^p \ z^q \, ,
\\
I^{p}(\xi) &=& \frac{1}{a^{p+1}} \int \dint z \ \dint \rho \ 
\frac{e^{-\sqrt{\rho^2+(z-\xi a)^2}/a} \ e^{-\sqrt{\rho^2+z^2}/a}}
                    {\sqrt{\rho^2+ (z-\xi a)^2} \sqrt{\rho^2+ z^2} }
                              \ \rho \ z^p \, .
\end{eqnarray}
\end{subequations}
Using the integrals  
$I^{3,1}(\xi)= \frac{\xi}{3} e^{-\xi}$,
$I^{1,3}(\xi) = \left(\frac{\xi}{2} + \frac{\xi^2}{3}\right) e^{-\xi}$, 
$I^{3,0}(\xi) = e^{-\xi}$, 
$I^{1,2}(\xi) = \frac{1}{2} (1+\xi) \, e^{-\xi}$, $I^{0}(\xi) = e^{-\xi}$, and $I^{1}(\xi) = \frac{\xi}{2} e^{-\xi}$,
we obtain for the cubic matrix element \eqref{eq:res_matrix_element_3}.

\section{Application of the semiclassical approach to zincblende semiconductors}
\label{sec:appendixB}

In the case of impurities in semiconductors with zincblende crystal structure, 
the spin-orbit coupling yields the symmetric form \cite{intronati-PRL2012,wellens-jalabert-2016}
\begin{subequations}
\begin{eqnarray}
\mathcal{C}_x(\mathbf{r})&=&-\frac{\gamma}{3a^5r}x\left(y^2-z^2\right)e^{-r/a}\\
\mathcal{C}_y(\mathbf{r})&=&-\frac{\gamma}{3a^5r}y\left(z^2-x^2\right)e^{-r/a}\\
\mathcal{C}_z(\mathbf{r})&=&-\frac{\gamma}{3a^5r}z\left(x^2-y^2\right)e^{-r/a} ,
\end{eqnarray}
\end{subequations}
where $\gamma$ is the Dresselhaus spin-orbit coupling strength. Using these 
expressions in \eqref{eq:reltimegen} and performing the spatial integral, one gets 
with \eqref{eq:hoppingtime} the final result for the spin-relaxation time in 
zincblende structures
\begin{equation}\label{eq:spinrelaxzb}
\left\langle \frac{1}{\tau^\mathrm{ZB}_\mathrm{s}} \right\rangle = \frac{8\sqrt{14\pi}}{147} 
\frac{\gamma^2}{a^6 V_0 \hbar} \sqrt{ n_\mathrm{i} a^3} ,
\end{equation}
where the value of the numerical prefactor is approximately $0.36$, correcting the 
one of Eq.\ (18) in Ref.\ \onlinecite{intronati-PRL2012}. 

\section{Fourier transform of the hopping amplitude matrix}
\label{app:ftham}

The diagrammatic expansions are more easily performed in reciprocal space, 
therefore it is useful to work with the Fourier transform of the hopping amplitude matrix
\begin{equation}
\label{eq:vtilde}
\tilde{\mathcal V}({\bf k})
=
\left(\begin{array}{cc} \tilde{\mathcal V}_0({\bf k}) + i \tilde{\mathcal C}_z({\bf k})& i \tilde{\mathcal C}_x({\bf k})+\tilde{\mathcal C}_y({\bf k}) \\
i \tilde{\mathcal C}_x({\bf k})-\tilde{\mathcal C}_y({\bf k}) & \tilde{\mathcal V}_0({\bf k}) -i \tilde{\mathcal C}_z({\bf k})\end{array}\right) \, .
\end{equation}

The spin-independent part is given by 
\begin{equation}
\tilde{\mathcal V}_{0}({\bf k}) =  -\frac{32 a^3 \pi V_0}{(1+a^2k^2)^3} \, .
\end{equation}
In the case of the WZ crystal structure we have
$\tilde{\mathcal{C}}_z(r)=0$, and thus
$\tilde{\mathcal{C}}_j(\mathbf{k})=\tilde{\mathcal{C}}^{(1)}_j(\mathbf{k})+
\tilde{\mathcal{C}}^{(3)}_j(\mathbf{k})$ for $j=x,y$.
These Fourier transforms can be readily done by a rotation of the integration variables
that places the new $z$-axis in the direction of the wave vector $\mathbf{k}$:
\begin{subequations}
\begin{eqnarray}
\tilde{\mathcal{C}}_x^{(1)}(\mathbf{k}) &=&  \frac{64 \alpha a^3 \pi i}{(1+(ka)^2)^4} k_y \, , \\
\tilde{\mathcal{C}}_y^{(1)}(\mathbf{k}) &=& -\frac{64 \alpha a^3 \pi i}{(1+(ka)^2)^4} k_x \,  ,
\end{eqnarray}
\end{subequations}
\begin{subequations}
\begin{eqnarray}
\tilde{\mathcal C}_x^{(3)}({\bf k}) & = &  -\frac{32 \pi  i a \gamma k_y \left[1+a^2\Bigl(7k^2-6(1+b)k_z^2\Bigr)\right]}{3(1+a^2k^2)^4} \, , \\
\tilde{\mathcal C}_y^{(3)}({\bf k}) & = & \frac{32 \pi  i a \gamma k_x \left[1+a^2\Bigl(7k^2-6(1+b)k_z^2\Bigr)\right]}{3(1+a^2k^2)^4} \, .
\end{eqnarray}
\label{eq:Cxyz}
\end{subequations}
These analytical expressions allow us to perform the relevant integrals of our work, 
like those of Eq.\ \eqref{eq:u2tilde}.

\bibliography{wurtzite_2016}

\begin{thebibliography}{65}%
\makeatletter
\providecommand \@ifxundefined [1]{%
 \@ifx{#1\undefined}
}%
\providecommand \@ifnum [1]{%
 \ifnum #1\expandafter \@firstoftwo
 \else \expandafter \@secondoftwo
 \fi
}%
\providecommand \@ifx [1]{%
 \ifx #1\expandafter \@firstoftwo
 \else \expandafter \@secondoftwo
 \fi
}%
\providecommand \natexlab [1]{#1}%
\providecommand \enquote  [1]{``#1''}%
\providecommand \bibnamefont  [1]{#1}%
\providecommand \bibfnamefont [1]{#1}%
\providecommand \citenamefont [1]{#1}%
\providecommand \href@noop [0]{\@secondoftwo}%
\providecommand \href [0]{\begingroup \@sanitize@url \@href}%
\providecommand \@href[1]{\@@startlink{#1}\@@href}%
\providecommand \@@href[1]{\endgroup#1\@@endlink}%
\providecommand \@sanitize@url [0]{\catcode `\\12\catcode `\$12\catcode
  `\&12\catcode `\#12\catcode `\^12\catcode `\_12\catcode `\%12\relax}%
\providecommand \@@startlink[1]{}%
\providecommand \@@endlink[0]{}%
\providecommand \url  [0]{\begingroup\@sanitize@url \@url }%
\providecommand \@url [1]{\endgroup\@href {#1}{\urlprefix }}%
\providecommand \urlprefix  [0]{URL }%
\providecommand \Eprint [0]{\href }%
\providecommand \doibase [0]{http://dx.doi.org/}%
\providecommand \selectlanguage [0]{\@gobble}%
\providecommand \bibinfo  [0]{\@secondoftwo}%
\providecommand \bibfield  [0]{\@secondoftwo}%
\providecommand \translation [1]{[#1]}%
\providecommand \BibitemOpen [0]{}%
\providecommand \bibitemStop [0]{}%
\providecommand \bibitemNoStop [0]{.\EOS\space}%
\providecommand \EOS [0]{\spacefactor3000\relax}%
\providecommand \BibitemShut  [1]{\csname bibitem#1\endcsname}%
\let\auto@bib@innerbib\@empty
\bibitem [{\citenamefont {Hanada}(2009)}]{Hanada2009}%
  \BibitemOpen
  \bibfield  {author} {\bibinfo {author} {\bibfnamefont {T.}~\bibnamefont
  {Hanada}},\ }\enquote {\bibinfo {title} {Basic properties of \mbox{ZnO},
  \mbox{GaN}, and related materials},}\ in\ \href {\doibase
  10.1007/978-3-540-88847-5_1} {\emph {\bibinfo {booktitle} {Oxide and Nitride
  Semiconductors: Processing, Properties, and Applications}}},\ \bibinfo
  {editor} {edited by\ \bibinfo {editor} {\bibfnamefont {T.}~\bibnamefont
  {Yao}}\ and\ \bibinfo {editor} {\bibfnamefont {S.-K.}\ \bibnamefont {Hong}}}\
  (\bibinfo  {publisher} {Springer Berlin Heidelberg},\ \bibinfo {address}
  {Berlin, Heidelberg},\ \bibinfo {year} {2009})\ pp.\ \bibinfo {pages}
  {1--19}\BibitemShut {NoStop}%
\bibitem [{\citenamefont {Dietl}\ \emph {et~al.}(2000)\citenamefont {Dietl},
  \citenamefont {Ohno}, \citenamefont {Matsukura}, \citenamefont {Cibert},\
  and\ \citenamefont {Ferrand}}]{Dietl1019}%
  \BibitemOpen
  \bibfield  {author} {\bibinfo {author} {\bibfnamefont {T.}~\bibnamefont
  {Dietl}}, \bibinfo {author} {\bibfnamefont {H.}~\bibnamefont {Ohno}},
  \bibinfo {author} {\bibfnamefont {F.}~\bibnamefont {Matsukura}}, \bibinfo
  {author} {\bibfnamefont {J.}~\bibnamefont {Cibert}}, \ and\ \bibinfo {author}
  {\bibfnamefont {D.}~\bibnamefont {Ferrand}},\ }\href {\doibase
  10.1126/science.287.5455.1019} {\bibfield  {journal} {\bibinfo  {journal}
  {Science}\ }\textbf {\bibinfo {volume} {287}},\ \bibinfo {pages} {1019}
  (\bibinfo {year} {2000})}\BibitemShut {NoStop}%
\bibitem [{\citenamefont {Pearton}\ \emph {et~al.}(2003)\citenamefont
  {Pearton}, \citenamefont {Abernathy}, \citenamefont {Overberg}, \citenamefont
  {Thaler}, \citenamefont {Norton}, \citenamefont {Theodoropoulou},
  \citenamefont {Hebard}, \citenamefont {Park}, \citenamefont {Ren},
  \citenamefont {Kim},\ and\ \citenamefont {Boatner}}]{pearton-JAP2003}%
  \BibitemOpen
  \bibfield  {author} {\bibinfo {author} {\bibfnamefont {S.~J.}\ \bibnamefont
  {Pearton}}, \bibinfo {author} {\bibfnamefont {C.~R.}\ \bibnamefont
  {Abernathy}}, \bibinfo {author} {\bibfnamefont {M.~E.}\ \bibnamefont
  {Overberg}}, \bibinfo {author} {\bibfnamefont {G.~T.}\ \bibnamefont
  {Thaler}}, \bibinfo {author} {\bibfnamefont {D.~P.}\ \bibnamefont {Norton}},
  \bibinfo {author} {\bibfnamefont {N.}~\bibnamefont {Theodoropoulou}},
  \bibinfo {author} {\bibfnamefont {A.~F.}\ \bibnamefont {Hebard}}, \bibinfo
  {author} {\bibfnamefont {Y.~D.}\ \bibnamefont {Park}}, \bibinfo {author}
  {\bibfnamefont {F.}~\bibnamefont {Ren}}, \bibinfo {author} {\bibfnamefont
  {J.}~\bibnamefont {Kim}}, \ and\ \bibinfo {author} {\bibfnamefont {L.~A.}\
  \bibnamefont {Boatner}},\ }\href {\doibase 10.1063/1.1517164} {\bibfield
  {journal} {\bibinfo  {journal} {J. Appl. Phys.}\ }\textbf {\bibinfo {volume}
  {93}},\ \bibinfo {pages} {1} (\bibinfo {year} {2003})}\BibitemShut {NoStop}%
\bibitem [{\citenamefont {\"Ozg\"ur}\ \emph {et~al.}(2005)\citenamefont
  {\"Ozg\"ur}, \citenamefont {Alivov}, \citenamefont {Liu}, \citenamefont
  {Teke}, \citenamefont {Reshchikov}, \citenamefont {Do\v{g}an}, \citenamefont
  {Avrutin}, \citenamefont {Cho},\ and\ \citenamefont
  {Morko\c{c}}}]{ozgur-JAP2005}%
  \BibitemOpen
  \bibfield  {author} {\bibinfo {author} {\bibfnamefont {U.}~\bibnamefont
  {\"Ozg\"ur}}, \bibinfo {author} {\bibfnamefont {Y.~I.}\ \bibnamefont
  {Alivov}}, \bibinfo {author} {\bibfnamefont {C.}~\bibnamefont {Liu}},
  \bibinfo {author} {\bibfnamefont {A.}~\bibnamefont {Teke}}, \bibinfo {author}
  {\bibfnamefont {M.~A.}\ \bibnamefont {Reshchikov}}, \bibinfo {author}
  {\bibfnamefont {S.}~\bibnamefont {Do\v{g}an}}, \bibinfo {author}
  {\bibfnamefont {V.}~\bibnamefont {Avrutin}}, \bibinfo {author} {\bibfnamefont
  {S.-J.}\ \bibnamefont {Cho}}, \ and\ \bibinfo {author} {\bibfnamefont
  {H.}~\bibnamefont {Morko\c{c}}},\ }\href {\doibase 10.1063/1.1992666}
  {\bibfield  {journal} {\bibinfo  {journal} {J. Appl. Phys.}\ }\textbf
  {\bibinfo {volume} {98}},\ \bibinfo {pages} {041301} (\bibinfo {year}
  {2005})}\BibitemShut {NoStop}%
\bibitem [{\citenamefont {Janotti}\ and\ \citenamefont
  {de~Walle}(2009)}]{janotti-RPP2009}%
  \BibitemOpen
  \bibfield  {author} {\bibinfo {author} {\bibfnamefont {A.}~\bibnamefont
  {Janotti}}\ and\ \bibinfo {author} {\bibfnamefont {C.~G.~V.}\ \bibnamefont
  {de~Walle}},\ }\href {http://stacks.iop.org/0034-4885/72/i=12/a=126501}
  {\bibfield  {journal} {\bibinfo  {journal} {Reports on Progress in Physics}\
  }\textbf {\bibinfo {volume} {72}},\ \bibinfo {pages} {126501} (\bibinfo
  {year} {2009})}\BibitemShut {NoStop}%
\bibitem [{\citenamefont {Krishnamurthy}\ \emph {et~al.}(2003)\citenamefont
  {Krishnamurthy}, \citenamefont {van Schilfgaarde},\ and\ \citenamefont
  {Newman}}]{kri-van-new}%
  \BibitemOpen
  \bibfield  {author} {\bibinfo {author} {\bibfnamefont {S.}~\bibnamefont
  {Krishnamurthy}}, \bibinfo {author} {\bibfnamefont {M.}~\bibnamefont {van
  Schilfgaarde}}, \ and\ \bibinfo {author} {\bibfnamefont {N.}~\bibnamefont
  {Newman}},\ }\href {\doibase 10.1063/1.1606873} {\bibfield  {journal}
  {\bibinfo  {journal} {Applied Physics Letters}\ }\textbf {\bibinfo {volume}
  {83}},\ \bibinfo {pages} {1761} (\bibinfo {year} {2003})}\BibitemShut
  {NoStop}%
\bibitem [{\citenamefont {Soumyanarayanan}\ \emph {et~al.}(2016)\citenamefont
  {Soumyanarayanan}, \citenamefont {Reyren}, \citenamefont {Fert},\ and\
  \citenamefont {Panagopoulos}}]{sou-rey-fer-pan}%
  \BibitemOpen
  \bibfield  {author} {\bibinfo {author} {\bibfnamefont {A.}~\bibnamefont
  {Soumyanarayanan}}, \bibinfo {author} {\bibfnamefont {N.}~\bibnamefont
  {Reyren}}, \bibinfo {author} {\bibfnamefont {A.}~\bibnamefont {Fert}}, \ and\
  \bibinfo {author} {\bibfnamefont {C.}~\bibnamefont {Panagopoulos}},\ }\href
  {\doibase 10.1038/nature19820} {\bibfield  {journal} {\bibinfo  {journal}
  {Nature}\ }\textbf {\bibinfo {volume} {539}},\ \bibinfo {pages} {509}
  (\bibinfo {year} {2016})}\BibitemShut {NoStop}%
\bibitem [{\citenamefont {Awschalom}\ \emph {et~al.}(2013)\citenamefont
  {Awschalom}, \citenamefont {Bassett}, \citenamefont {Dzurak}, \citenamefont
  {Hu},\ and\ \citenamefont {Petta}}]{aws-bas-dzu}%
  \BibitemOpen
  \bibfield  {author} {\bibinfo {author} {\bibfnamefont {D.~D.}\ \bibnamefont
  {Awschalom}}, \bibinfo {author} {\bibfnamefont {L.~C.}\ \bibnamefont
  {Bassett}}, \bibinfo {author} {\bibfnamefont {A.~S.}\ \bibnamefont {Dzurak}},
  \bibinfo {author} {\bibfnamefont {E.~L.}\ \bibnamefont {Hu}}, \ and\ \bibinfo
  {author} {\bibfnamefont {J.~R.}\ \bibnamefont {Petta}},\ }\href {\doibase
  10.1126/science.1231364} {\bibfield  {journal} {\bibinfo  {journal}
  {Science}\ }\textbf {\bibinfo {volume} {339}},\ \bibinfo {pages} {1174}
  (\bibinfo {year} {2013})}\BibitemShut {NoStop}%
\bibitem [{\citenamefont {Bader}\ and\ \citenamefont {Parkin}(2010)}]{bad-par}%
  \BibitemOpen
  \bibfield  {author} {\bibinfo {author} {\bibfnamefont {S.}~\bibnamefont
  {Bader}}\ and\ \bibinfo {author} {\bibfnamefont {S.}~\bibnamefont {Parkin}},\
  }\href {\doibase 10.1146/annurev-conmatphys-070909-104123} {\bibfield
  {journal} {\bibinfo  {journal} {Annual Review of Condensed Matter Physics}\
  }\textbf {\bibinfo {volume} {1}},\ \bibinfo {pages} {71} (\bibinfo {year}
  {2010})}\BibitemShut {NoStop}%
\bibitem [{\citenamefont {Fabian}\ \emph {et~al.}(2007)\citenamefont {Fabian},
  \citenamefont {Matos-Abiague}, \citenamefont {Ertler}, \citenamefont
  {Stano},\ and\ \citenamefont {\v{Z}uti\'c}}]{fab-mat-ert}%
  \BibitemOpen
  \bibfield  {author} {\bibinfo {author} {\bibfnamefont {J.}~\bibnamefont
  {Fabian}}, \bibinfo {author} {\bibfnamefont {A.}~\bibnamefont
  {Matos-Abiague}}, \bibinfo {author} {\bibfnamefont {C.}~\bibnamefont
  {Ertler}}, \bibinfo {author} {\bibfnamefont {P.}~\bibnamefont {Stano}}, \
  and\ \bibinfo {author} {\bibfnamefont {I.}~\bibnamefont {\v{Z}uti\'c}},\
  }\href {http://www.physics.sk/aps/pub.php?y=2007&pub=aps-07-04} {\bibfield
  {journal} {\bibinfo  {journal} {Acta Physica Slovaca}\ }\textbf {\bibinfo
  {volume} {57}},\ \bibinfo {pages} {565} (\bibinfo {year} {2007})}\BibitemShut
  {NoStop}%
\bibitem [{\citenamefont {Awschalom}\ \emph {et~al.}(2002)\citenamefont
  {Awschalom}, \citenamefont {Loss},\ and\ \citenamefont
  {Samarth}}]{awschalom-book2002}%
  \BibitemOpen
  \bibinfo {editor} {\bibfnamefont {D.}~\bibnamefont {Awschalom}}, \bibinfo
  {editor} {\bibfnamefont {D.}~\bibnamefont {Loss}}, \ and\ \bibinfo {editor}
  {\bibfnamefont {N.}~\bibnamefont {Samarth}},\ eds.,\ in\ \href
  {http://www.springer.com/in/book/9783540421764} {\emph {\bibinfo {booktitle}
  {Semiconductor Spintronics and Quantum Computation}}}\ (\bibinfo  {publisher}
  {Springer Berlin Heidelberg},\ \bibinfo {address} {Berlin, Heidelberg},\
  \bibinfo {year} {2002})\BibitemShut {NoStop}%
\bibitem [{\citenamefont {\ifmmode \check{Z}\else
  \v{Z}\fi{}uti\ifmmode~\acute{c}\else \'{c}\fi{}}\ \emph
  {et~al.}(2004)\citenamefont {\ifmmode \check{Z}\else
  \v{Z}\fi{}uti\ifmmode~\acute{c}\else \'{c}\fi{}}, \citenamefont {Fabian},\
  and\ \citenamefont {Das~Sarma}}]{zut-fab-das}%
  \BibitemOpen
  \bibfield  {author} {\bibinfo {author} {\bibfnamefont {I.}~\bibnamefont
  {\ifmmode \check{Z}\else \v{Z}\fi{}uti\ifmmode~\acute{c}\else \'{c}\fi{}}},
  \bibinfo {author} {\bibfnamefont {J.}~\bibnamefont {Fabian}}, \ and\ \bibinfo
  {author} {\bibfnamefont {S.}~\bibnamefont {Das~Sarma}},\ }\href {\doibase
  10.1103/RevModPhys.76.323} {\bibfield  {journal} {\bibinfo  {journal} {Rev.
  Mod. Phys.}\ }\textbf {\bibinfo {volume} {76}},\ \bibinfo {pages} {323}
  (\bibinfo {year} {2004})}\BibitemShut {NoStop}%
\bibitem [{\citenamefont {Kikkawa}\ and\ \citenamefont
  {Awschalom}(1998)}]{kik-aws}%
  \BibitemOpen
  \bibfield  {author} {\bibinfo {author} {\bibfnamefont {J.~M.}\ \bibnamefont
  {Kikkawa}}\ and\ \bibinfo {author} {\bibfnamefont {D.~D.}\ \bibnamefont
  {Awschalom}},\ }\href {\doibase 10.1103/PhysRevLett.80.4313} {\bibfield
  {journal} {\bibinfo  {journal} {Phys. Rev. Lett.}\ }\textbf {\bibinfo
  {volume} {80}},\ \bibinfo {pages} {4313} (\bibinfo {year}
  {1998})}\BibitemShut {NoStop}%
\bibitem [{\citenamefont {Dzhioev}\ \emph {et~al.}(2002)\citenamefont
  {Dzhioev}, \citenamefont {Kavokin}, \citenamefont {Korenev}, \citenamefont
  {Lazarev}, \citenamefont {Meltser}, \citenamefont {Stepanova}, \citenamefont
  {Zakharchenya}, \citenamefont {Gammon},\ and\ \citenamefont
  {Katzer}}]{dzh-kav-kor}%
  \BibitemOpen
  \bibfield  {author} {\bibinfo {author} {\bibfnamefont {R.~I.}\ \bibnamefont
  {Dzhioev}}, \bibinfo {author} {\bibfnamefont {K.~V.}\ \bibnamefont
  {Kavokin}}, \bibinfo {author} {\bibfnamefont {V.~L.}\ \bibnamefont
  {Korenev}}, \bibinfo {author} {\bibfnamefont {M.~V.}\ \bibnamefont
  {Lazarev}}, \bibinfo {author} {\bibfnamefont {B.~Y.}\ \bibnamefont
  {Meltser}}, \bibinfo {author} {\bibfnamefont {M.~N.}\ \bibnamefont
  {Stepanova}}, \bibinfo {author} {\bibfnamefont {B.~P.}\ \bibnamefont
  {Zakharchenya}}, \bibinfo {author} {\bibfnamefont {D.}~\bibnamefont
  {Gammon}}, \ and\ \bibinfo {author} {\bibfnamefont {D.~S.}\ \bibnamefont
  {Katzer}},\ }\href {\doibase 10.1103/PhysRevB.66.245204} {\bibfield
  {journal} {\bibinfo  {journal} {Phys. Rev. B}\ }\textbf {\bibinfo {volume}
  {66}},\ \bibinfo {pages} {245204} (\bibinfo {year} {2002})}\BibitemShut
  {NoStop}%
\bibitem [{\citenamefont {Oestreich}\ \emph {et~al.}(2005)\citenamefont
  {Oestreich}, \citenamefont {R\"omer}, \citenamefont {Haug},\ and\
  \citenamefont {H\"agele}}]{oestreich-PRL2005}%
  \BibitemOpen
  \bibfield  {author} {\bibinfo {author} {\bibfnamefont {M.}~\bibnamefont
  {Oestreich}}, \bibinfo {author} {\bibfnamefont {M.}~\bibnamefont {R\"omer}},
  \bibinfo {author} {\bibfnamefont {R.~J.}\ \bibnamefont {Haug}}, \ and\
  \bibinfo {author} {\bibfnamefont {D.}~\bibnamefont {H\"agele}},\ }\href
  {\doibase 10.1103/PhysRevLett.95.216603} {\bibfield  {journal} {\bibinfo
  {journal} {Phys. Rev. Lett.}\ }\textbf {\bibinfo {volume} {95}},\ \bibinfo
  {pages} {216603} (\bibinfo {year} {2005})}\BibitemShut {NoStop}%
\bibitem [{\citenamefont {Wu}\ \emph {et~al.}(2010)\citenamefont {Wu},
  \citenamefont {Jiang},\ and\ \citenamefont {Weng}}]{wu-jia-wen}%
  \BibitemOpen
  \bibfield  {author} {\bibinfo {author} {\bibfnamefont {M.}~\bibnamefont
  {Wu}}, \bibinfo {author} {\bibfnamefont {J.}~\bibnamefont {Jiang}}, \ and\
  \bibinfo {author} {\bibfnamefont {M.}~\bibnamefont {Weng}},\ }\href {\doibase
  http://dx.doi.org/10.1016/j.physrep.2010.04.002} {\bibfield  {journal}
  {\bibinfo  {journal} {Phys. Reports}\ }\textbf {\bibinfo {volume} {493}},\
  \bibinfo {pages} {61} (\bibinfo {year} {2010})}\BibitemShut {NoStop}%
\bibitem [{\citenamefont {Tamborenea}\ \emph {et~al.}(2007)\citenamefont
  {Tamborenea}, \citenamefont {Weinmann},\ and\ \citenamefont
  {Jalabert}}]{tamborenea-PRB2007}%
  \BibitemOpen
  \bibfield  {author} {\bibinfo {author} {\bibfnamefont {P.~I.}\ \bibnamefont
  {Tamborenea}}, \bibinfo {author} {\bibfnamefont {D.}~\bibnamefont
  {Weinmann}}, \ and\ \bibinfo {author} {\bibfnamefont {R.~A.}\ \bibnamefont
  {Jalabert}},\ }\href {\doibase 10.1103/PhysRevB.76.085209} {\bibfield
  {journal} {\bibinfo  {journal} {Phys. Rev. B}\ }\textbf {\bibinfo {volume}
  {76}},\ \bibinfo {pages} {085209} (\bibinfo {year} {2007})}\BibitemShut
  {NoStop}%
\bibitem [{\citenamefont {Matsubara}\ and\ \citenamefont
  {Toyozawa}(1961)}]{matsubara-1961}%
  \BibitemOpen
  \bibfield  {author} {\bibinfo {author} {\bibfnamefont {T.}~\bibnamefont
  {Matsubara}}\ and\ \bibinfo {author} {\bibfnamefont {Y.}~\bibnamefont
  {Toyozawa}},\ }\href {\doibase 10.1143/PTP.26.739} {\bibfield  {journal}
  {\bibinfo  {journal} {Progress of Theoretical Physics}\ }\textbf {\bibinfo
  {volume} {26}},\ \bibinfo {pages} {739} (\bibinfo {year} {1961})}\BibitemShut
  {NoStop}%
\bibitem [{\citenamefont {Intronati}\ \emph {et~al.}(2012)\citenamefont
  {Intronati}, \citenamefont {Tamborenea}, \citenamefont {Weinmann},\ and\
  \citenamefont {Jalabert}}]{intronati-PRL2012}%
  \BibitemOpen
  \bibfield  {author} {\bibinfo {author} {\bibfnamefont {G.~A.}\ \bibnamefont
  {Intronati}}, \bibinfo {author} {\bibfnamefont {P.~I.}\ \bibnamefont
  {Tamborenea}}, \bibinfo {author} {\bibfnamefont {D.}~\bibnamefont
  {Weinmann}}, \ and\ \bibinfo {author} {\bibfnamefont {R.~A.}\ \bibnamefont
  {Jalabert}},\ }\href {\doibase 10.1103/PhysRevLett.108.016601} {\bibfield
  {journal} {\bibinfo  {journal} {Phys. Rev. Lett.}\ }\textbf {\bibinfo
  {volume} {108}},\ \bibinfo {pages} {016601} (\bibinfo {year}
  {2012})}\BibitemShut {NoStop}%
\bibitem [{\citenamefont {Wellens}\ and\ \citenamefont
  {Jalabert}(2016)}]{wellens-jalabert-2016}%
  \BibitemOpen
  \bibfield  {author} {\bibinfo {author} {\bibfnamefont {T.}~\bibnamefont
  {Wellens}}\ and\ \bibinfo {author} {\bibfnamefont {R.~A.}\ \bibnamefont
  {Jalabert}},\ }\href {\doibase 10.1103/PhysRevB.94.144209} {\bibfield
  {journal} {\bibinfo  {journal} {Phys. Rev. B}\ }\textbf {\bibinfo {volume}
  {94}},\ \bibinfo {pages} {144209} (\bibinfo {year} {2016})}\BibitemShut
  {NoStop}%
\bibitem [{\citenamefont {Beschoten}\ \emph {et~al.}(2001)\citenamefont
  {Beschoten}, \citenamefont {Johnston-Halperin}, \citenamefont {Young},
  \citenamefont {Poggio}, \citenamefont {Grimaldi}, \citenamefont {Keller},
  \citenamefont {DenBaars}, \citenamefont {Mishra}, \citenamefont {Hu},\ and\
  \citenamefont {Awschalom}}]{bes-etal}%
  \BibitemOpen
  \bibfield  {author} {\bibinfo {author} {\bibfnamefont {B.}~\bibnamefont
  {Beschoten}}, \bibinfo {author} {\bibfnamefont {E.}~\bibnamefont
  {Johnston-Halperin}}, \bibinfo {author} {\bibfnamefont {D.~K.}\ \bibnamefont
  {Young}}, \bibinfo {author} {\bibfnamefont {M.}~\bibnamefont {Poggio}},
  \bibinfo {author} {\bibfnamefont {J.~E.}\ \bibnamefont {Grimaldi}}, \bibinfo
  {author} {\bibfnamefont {S.}~\bibnamefont {Keller}}, \bibinfo {author}
  {\bibfnamefont {S.~P.}\ \bibnamefont {DenBaars}}, \bibinfo {author}
  {\bibfnamefont {U.~K.}\ \bibnamefont {Mishra}}, \bibinfo {author}
  {\bibfnamefont {E.~L.}\ \bibnamefont {Hu}}, \ and\ \bibinfo {author}
  {\bibfnamefont {D.~D.}\ \bibnamefont {Awschalom}},\ }\href {\doibase
  10.1103/PhysRevB.63.121202} {\bibfield  {journal} {\bibinfo  {journal} {Phys.
  Rev. B}\ }\textbf {\bibinfo {volume} {63}},\ \bibinfo {pages} {121202}
  (\bibinfo {year} {2001})}\BibitemShut {NoStop}%
\bibitem [{\citenamefont {Wolos}\ \emph
  {et~al.}(2011{\natexlab{a}})\citenamefont {Wolos}, \citenamefont
  {Wilamowski}, \citenamefont {Piersa}, \citenamefont {Strupinski},
  \citenamefont {Lucznik}, \citenamefont {Grzegory},\ and\ \citenamefont
  {Porowski}}]{wol-etal}%
  \BibitemOpen
  \bibfield  {author} {\bibinfo {author} {\bibfnamefont {A.}~\bibnamefont
  {Wolos}}, \bibinfo {author} {\bibfnamefont {Z.}~\bibnamefont {Wilamowski}},
  \bibinfo {author} {\bibfnamefont {M.}~\bibnamefont {Piersa}}, \bibinfo
  {author} {\bibfnamefont {W.}~\bibnamefont {Strupinski}}, \bibinfo {author}
  {\bibfnamefont {B.}~\bibnamefont {Lucznik}}, \bibinfo {author} {\bibfnamefont
  {I.}~\bibnamefont {Grzegory}}, \ and\ \bibinfo {author} {\bibfnamefont
  {S.}~\bibnamefont {Porowski}},\ }\href {\doibase 10.1103/PhysRevB.83.165206}
  {\bibfield  {journal} {\bibinfo  {journal} {Phys. Rev. B}\ }\textbf {\bibinfo
  {volume} {83}},\ \bibinfo {pages} {165206} (\bibinfo {year}
  {2011}{\natexlab{a}})}\BibitemShut {NoStop}%
\bibitem [{\citenamefont {Bu\ss{}}\ \emph {et~al.}(2010)\citenamefont
  {Bu\ss{}}, \citenamefont {Rudolph}, \citenamefont {Natali}, \citenamefont
  {Semond},\ and\ \citenamefont {H\"agele}}]{bus-etal-2010a}%
  \BibitemOpen
  \bibfield  {author} {\bibinfo {author} {\bibfnamefont {J.~H.}\ \bibnamefont
  {Bu\ss{}}}, \bibinfo {author} {\bibfnamefont {J.}~\bibnamefont {Rudolph}},
  \bibinfo {author} {\bibfnamefont {F.}~\bibnamefont {Natali}}, \bibinfo
  {author} {\bibfnamefont {F.}~\bibnamefont {Semond}}, \ and\ \bibinfo {author}
  {\bibfnamefont {D.}~\bibnamefont {H\"agele}},\ }\href {\doibase
  10.1103/PhysRevB.81.155216} {\bibfield  {journal} {\bibinfo  {journal} {Phys.
  Rev. B}\ }\textbf {\bibinfo {volume} {81}},\ \bibinfo {pages} {155216}
  (\bibinfo {year} {2010})}\BibitemShut {NoStop}%
\bibitem [{\citenamefont {Bu\ss{}}\ \emph {et~al.}(2011)\citenamefont
  {Bu\ss{}}, \citenamefont {Rudolph}, \citenamefont {Starosielec},
  \citenamefont {Schaefer}, \citenamefont {Semond}, \citenamefont {Cordier},
  \citenamefont {Wieck},\ and\ \citenamefont {Hagele}}]{bus-etal-2011}%
  \BibitemOpen
  \bibfield  {author} {\bibinfo {author} {\bibfnamefont {J.~H.}\ \bibnamefont
  {Bu\ss{}}}, \bibinfo {author} {\bibfnamefont {J.}~\bibnamefont {Rudolph}},
  \bibinfo {author} {\bibfnamefont {S.}~\bibnamefont {Starosielec}}, \bibinfo
  {author} {\bibfnamefont {A.}~\bibnamefont {Schaefer}}, \bibinfo {author}
  {\bibfnamefont {F.}~\bibnamefont {Semond}}, \bibinfo {author} {\bibfnamefont
  {Y.}~\bibnamefont {Cordier}}, \bibinfo {author} {\bibfnamefont {A.~D.}\
  \bibnamefont {Wieck}}, \ and\ \bibinfo {author} {\bibfnamefont
  {D.}~\bibnamefont {Hagele}},\ }\href {\doibase 10.1103/PhysRevB.84.153202}
  {\bibfield  {journal} {\bibinfo  {journal} {Phys. Rev. B}\ }\textbf {\bibinfo
  {volume} {84}},\ \bibinfo {pages} {153202} (\bibinfo {year}
  {2011})}\BibitemShut {NoStop}%
\bibitem [{\citenamefont {Ghosh}\ \emph {et~al.}(2005)\citenamefont {Ghosh},
  \citenamefont {Sih}, \citenamefont {Lau}, \citenamefont {Awschalom},
  \citenamefont {Bae}, \citenamefont {Wang}, \citenamefont {Vaidya},\ and\
  \citenamefont {Chapline}}]{gho-etal}%
  \BibitemOpen
  \bibfield  {author} {\bibinfo {author} {\bibfnamefont {S.}~\bibnamefont
  {Ghosh}}, \bibinfo {author} {\bibfnamefont {V.}~\bibnamefont {Sih}}, \bibinfo
  {author} {\bibfnamefont {W.~H.}\ \bibnamefont {Lau}}, \bibinfo {author}
  {\bibfnamefont {D.~D.}\ \bibnamefont {Awschalom}}, \bibinfo {author}
  {\bibfnamefont {S.-Y.}\ \bibnamefont {Bae}}, \bibinfo {author} {\bibfnamefont
  {S.}~\bibnamefont {Wang}}, \bibinfo {author} {\bibfnamefont {S.}~\bibnamefont
  {Vaidya}}, \ and\ \bibinfo {author} {\bibfnamefont {G.}~\bibnamefont
  {Chapline}},\ }\href {\doibase 10.1063/1.1946204} {\bibfield  {journal}
  {\bibinfo  {journal} {Appl. Phys. Lett.}\ }\textbf {\bibinfo {volume} {86}},\
  \bibinfo {pages} {232507} (\bibinfo {year} {2005})}\BibitemShut {NoStop}%
\bibitem [{\citenamefont {Prestgard}\ \emph {et~al.}(2015)\citenamefont
  {Prestgard}, \citenamefont {Siegel}, \citenamefont {Roundy}, \citenamefont
  {Raikh},\ and\ \citenamefont {Tiwari}}]{pre-etal}%
  \BibitemOpen
  \bibfield  {author} {\bibinfo {author} {\bibfnamefont {M.~C.}\ \bibnamefont
  {Prestgard}}, \bibinfo {author} {\bibfnamefont {G.}~\bibnamefont {Siegel}},
  \bibinfo {author} {\bibfnamefont {R.}~\bibnamefont {Roundy}}, \bibinfo
  {author} {\bibfnamefont {M.}~\bibnamefont {Raikh}}, \ and\ \bibinfo {author}
  {\bibfnamefont {A.}~\bibnamefont {Tiwari}},\ }\href {\doibase
  10.1063/1.4913287} {\bibfield  {journal} {\bibinfo  {journal} {J. Appl.
  Phys.}\ }\textbf {\bibinfo {volume} {117}},\ \bibinfo {pages} {083905}
  (\bibinfo {year} {2015})}\BibitemShut {NoStop}%
\bibitem [{\citenamefont {Ghosh}\ \emph {et~al.}(2008)\citenamefont {Ghosh},
  \citenamefont {Steuerman}, \citenamefont {Maertz}, \citenamefont {Ohtani},
  \citenamefont {Xu}, \citenamefont {Ohno},\ and\ \citenamefont
  {Awschalom}}]{ghosh-APL2008}%
  \BibitemOpen
  \bibfield  {author} {\bibinfo {author} {\bibfnamefont {S.}~\bibnamefont
  {Ghosh}}, \bibinfo {author} {\bibfnamefont {D.~W.}\ \bibnamefont
  {Steuerman}}, \bibinfo {author} {\bibfnamefont {B.}~\bibnamefont {Maertz}},
  \bibinfo {author} {\bibfnamefont {K.}~\bibnamefont {Ohtani}}, \bibinfo
  {author} {\bibfnamefont {H.}~\bibnamefont {Xu}}, \bibinfo {author}
  {\bibfnamefont {H.}~\bibnamefont {Ohno}}, \ and\ \bibinfo {author}
  {\bibfnamefont {D.~D.}\ \bibnamefont {Awschalom}},\ }\href {\doibase
  10.1063/1.2913049} {\bibfield  {journal} {\bibinfo  {journal} {Applied
  Physics Letters}\ }\textbf {\bibinfo {volume} {92}},\ \bibinfo {pages}
  {162109} (\bibinfo {year} {2008})}\BibitemShut {NoStop}%
\bibitem [{\citenamefont {Ganichev}\ and\ \citenamefont
  {Golub}(2014)}]{ganichev-pss2014}%
  \BibitemOpen
  \bibfield  {author} {\bibinfo {author} {\bibfnamefont {S.~D.}\ \bibnamefont
  {Ganichev}}\ and\ \bibinfo {author} {\bibfnamefont {L.~E.}\ \bibnamefont
  {Golub}},\ }\href {\doibase 10.1002/pssb.201350261} {\bibfield  {journal}
  {\bibinfo  {journal} {physica status solidi (b)}\ }\textbf {\bibinfo {volume}
  {251}},\ \bibinfo {pages} {1801} (\bibinfo {year} {2014})}\BibitemShut
  {NoStop}%
\bibitem [{\citenamefont {Harmon}\ \emph {et~al.}(2011)\citenamefont {Harmon},
  \citenamefont {Putikka},\ and\ \citenamefont {Joynt}}]{harmon-APL2011}%
  \BibitemOpen
  \bibfield  {author} {\bibinfo {author} {\bibfnamefont {N.~J.}\ \bibnamefont
  {Harmon}}, \bibinfo {author} {\bibfnamefont {W.~O.}\ \bibnamefont {Putikka}},
  \ and\ \bibinfo {author} {\bibfnamefont {R.}~\bibnamefont {Joynt}},\ }\href
  {\doibase 10.1063/1.3555628} {\bibfield  {journal} {\bibinfo  {journal}
  {Appl. Phys. Lett.}\ }\textbf {\bibinfo {volume} {98}},\ \bibinfo {pages}
  {073108} (\bibinfo {year} {2011})}\BibitemShut {NoStop}%
\bibitem [{\citenamefont {Harmon}\ \emph {et~al.}(2009)\citenamefont {Harmon},
  \citenamefont {Putikka},\ and\ \citenamefont {Joynt}}]{harmon-PRB2009}%
  \BibitemOpen
  \bibfield  {author} {\bibinfo {author} {\bibfnamefont {N.~J.}\ \bibnamefont
  {Harmon}}, \bibinfo {author} {\bibfnamefont {W.~O.}\ \bibnamefont {Putikka}},
  \ and\ \bibinfo {author} {\bibfnamefont {R.}~\bibnamefont {Joynt}},\ }\href
  {\doibase 10.1103/PhysRevB.79.115204} {\bibfield  {journal} {\bibinfo
  {journal} {Phys. Rev. B}\ }\textbf {\bibinfo {volume} {79}},\ \bibinfo
  {pages} {115204} (\bibinfo {year} {2009})}\BibitemShut {NoStop}%
\bibitem [{\citenamefont {Wang}\ \emph {et~al.}(2010)\citenamefont {Wang},
  \citenamefont {Wu}, \citenamefont {Chiang}, \citenamefont {Lo}, \citenamefont
  {Kao}, \citenamefont {Hsu}, \citenamefont {Pang}, \citenamefont {Jang},
  \citenamefont {Lee}, \citenamefont {Chang},\ and\ \citenamefont
  {Chen}}]{wang-JAP2010}%
  \BibitemOpen
  \bibfield  {author} {\bibinfo {author} {\bibfnamefont {W.-T.}\ \bibnamefont
  {Wang}}, \bibinfo {author} {\bibfnamefont {C.~L.}\ \bibnamefont {Wu}},
  \bibinfo {author} {\bibfnamefont {J.~C.}\ \bibnamefont {Chiang}}, \bibinfo
  {author} {\bibfnamefont {I.}~\bibnamefont {Lo}}, \bibinfo {author}
  {\bibfnamefont {H.~F.}\ \bibnamefont {Kao}}, \bibinfo {author} {\bibfnamefont
  {Y.~C.}\ \bibnamefont {Hsu}}, \bibinfo {author} {\bibfnamefont {W.~Y.}\
  \bibnamefont {Pang}}, \bibinfo {author} {\bibfnamefont {D.~J.}\ \bibnamefont
  {Jang}}, \bibinfo {author} {\bibfnamefont {M.-E.}\ \bibnamefont {Lee}},
  \bibinfo {author} {\bibfnamefont {Y.-C.}\ \bibnamefont {Chang}}, \ and\
  \bibinfo {author} {\bibfnamefont {C.-N.}\ \bibnamefont {Chen}},\ }\href
  {\doibase 10.1063/1.3484042} {\bibfield  {journal} {\bibinfo  {journal} {J.
  Appl. Phys.}\ }\textbf {\bibinfo {volume} {108}},\ \bibinfo {pages} {083718}
  (\bibinfo {year} {2010})}\BibitemShut {NoStop}%
\bibitem [{\citenamefont {Kang}(2015)}]{kang-ps2015}%
  \BibitemOpen
  \bibfield  {author} {\bibinfo {author} {\bibfnamefont {N.~L.}\ \bibnamefont
  {Kang}},\ }\href {\doibase doi:10.1088/0031-8949/90/3/035805} {\bibfield
  {journal} {\bibinfo  {journal} {Physica Scripta}\ }\textbf {\bibinfo {volume}
  {90}},\ \bibinfo {pages} {035805} (\bibinfo {year} {2015})}\BibitemShut
  {NoStop}%
\bibitem [{\citenamefont {Bu\ss{}}\ \emph {et~al.}(2009)\citenamefont
  {Bu\ss{}}, \citenamefont {Rudolph}, \citenamefont {Natali}, \citenamefont
  {Semond},\ and\ \citenamefont {H\"agele}}]{bus-etal-2009}%
  \BibitemOpen
  \bibfield  {author} {\bibinfo {author} {\bibfnamefont {J.~H.}\ \bibnamefont
  {Bu\ss{}}}, \bibinfo {author} {\bibfnamefont {J.}~\bibnamefont {Rudolph}},
  \bibinfo {author} {\bibfnamefont {F.}~\bibnamefont {Natali}}, \bibinfo
  {author} {\bibfnamefont {F.}~\bibnamefont {Semond}}, \ and\ \bibinfo {author}
  {\bibfnamefont {D.}~\bibnamefont {H\"agele}},\ }\href {\doibase
  10.1063/1.3261755} {\bibfield  {journal} {\bibinfo  {journal} {Appl. Phys.
  Lett.}\ }\textbf {\bibinfo {volume} {95}},\ \bibinfo {pages} {192107}
  (\bibinfo {year} {2009})}\BibitemShut {NoStop}%
\bibitem [{\citenamefont {Cartoix\`a}\ \emph {et~al.}(2005)\citenamefont
  {Cartoix\`a}, \citenamefont {Ting},\ and\ \citenamefont
  {Chang}}]{cartoixa-PRB2005}%
  \BibitemOpen
  \bibfield  {author} {\bibinfo {author} {\bibfnamefont {X.}~\bibnamefont
  {Cartoix\`a}}, \bibinfo {author} {\bibfnamefont {D.~Z.-Y.}\ \bibnamefont
  {Ting}}, \ and\ \bibinfo {author} {\bibfnamefont {Y.-C.}\ \bibnamefont
  {Chang}},\ }\href {\doibase 10.1103/PhysRevB.71.045313} {\bibfield  {journal}
  {\bibinfo  {journal} {Phys. Rev. B}\ }\textbf {\bibinfo {volume} {71}},\
  \bibinfo {pages} {045313} (\bibinfo {year} {2005})}\BibitemShut {NoStop}%
\bibitem [{\citenamefont {L\"u}\ and\ \citenamefont
  {Cheng}(2009)}]{lu-SST2009}%
  \BibitemOpen
  \bibfield  {author} {\bibinfo {author} {\bibfnamefont {C.}~\bibnamefont
  {L\"u}}\ and\ \bibinfo {author} {\bibfnamefont {J.~L.}\ \bibnamefont
  {Cheng}},\ }\href {\doibase 10.1088/0268-1242/24/11/115010} {\bibfield
  {journal} {\bibinfo  {journal} {Semiconductor Science and Technology}\
  }\textbf {\bibinfo {volume} {24}},\ \bibinfo {pages} {115010} (\bibinfo
  {year} {2009})}\BibitemShut {NoStop}%
\bibitem [{\citenamefont {Fu}\ and\ \citenamefont {Wu}(2008)}]{fu-wu}%
  \BibitemOpen
  \bibfield  {author} {\bibinfo {author} {\bibfnamefont {J.~Y.}\ \bibnamefont
  {Fu}}\ and\ \bibinfo {author} {\bibfnamefont {M.~W.}\ \bibnamefont {Wu}},\
  }\href {\doibase 10.1063/1.3018600} {\bibfield  {journal} {\bibinfo
  {journal} {J. Appl. Phys.}\ }\textbf {\bibinfo {volume} {104}},\ \bibinfo
  {pages} {093712} (\bibinfo {year} {2008})}\BibitemShut {NoStop}%
\bibitem [{\citenamefont {Wang}\ \emph {et~al.}(2007)\citenamefont {Wang},
  \citenamefont {Wu}, \citenamefont {Tsay}, \citenamefont {Gau}, \citenamefont
  {Lo}, \citenamefont {Kao}, \citenamefont {Jang}, \citenamefont {Chiang},
  \citenamefont {Lee}, \citenamefont {Chang}, \citenamefont {Chen},\ and\
  \citenamefont {Hsueh}}]{wang-etal}%
  \BibitemOpen
  \bibfield  {author} {\bibinfo {author} {\bibfnamefont {W.-T.}\ \bibnamefont
  {Wang}}, \bibinfo {author} {\bibfnamefont {C.~L.}\ \bibnamefont {Wu}},
  \bibinfo {author} {\bibfnamefont {S.~F.}\ \bibnamefont {Tsay}}, \bibinfo
  {author} {\bibfnamefont {M.~H.}\ \bibnamefont {Gau}}, \bibinfo {author}
  {\bibfnamefont {I.}~\bibnamefont {Lo}}, \bibinfo {author} {\bibfnamefont
  {H.~F.}\ \bibnamefont {Kao}}, \bibinfo {author} {\bibfnamefont {D.~J.}\
  \bibnamefont {Jang}}, \bibinfo {author} {\bibfnamefont {J.-C.}\ \bibnamefont
  {Chiang}}, \bibinfo {author} {\bibfnamefont {M.-E.}\ \bibnamefont {Lee}},
  \bibinfo {author} {\bibfnamefont {Y.-C.}\ \bibnamefont {Chang}}, \bibinfo
  {author} {\bibfnamefont {C.-N.}\ \bibnamefont {Chen}}, \ and\ \bibinfo
  {author} {\bibfnamefont {H.~C.}\ \bibnamefont {Hsueh}},\ }\href {\doibase
  10.1063/1.2775038} {\bibfield  {journal} {\bibinfo  {journal} {Appl. Phys.
  Lett.}\ }\textbf {\bibinfo {volume} {91}},\ \bibinfo {pages} {082110}
  (\bibinfo {year} {2007})}\BibitemShut {NoStop}%
\bibitem [{\citenamefont {Xu}\ and\ \citenamefont {Ching}(1993)}]{xu-PRB1993}%
  \BibitemOpen
  \bibfield  {author} {\bibinfo {author} {\bibfnamefont {Y.-N.}\ \bibnamefont
  {Xu}}\ and\ \bibinfo {author} {\bibfnamefont {W.~Y.}\ \bibnamefont {Ching}},\
  }\href {\doibase 10.1103/PhysRevB.48.4335} {\bibfield  {journal} {\bibinfo
  {journal} {Phys. Rev. B}\ }\textbf {\bibinfo {volume} {48}},\ \bibinfo
  {pages} {4335} (\bibinfo {year} {1993})}\BibitemShut {NoStop}%
\bibitem [{\citenamefont {De}\ and\ \citenamefont {Pryor}(2010)}]{de-pry}%
  \BibitemOpen
  \bibfield  {author} {\bibinfo {author} {\bibfnamefont {A.}~\bibnamefont
  {De}}\ and\ \bibinfo {author} {\bibfnamefont {C.~E.}\ \bibnamefont {Pryor}},\
  }\href {\doibase 10.1103/PhysRevB.81.155210} {\bibfield  {journal} {\bibinfo
  {journal} {Phys. Rev. B}\ }\textbf {\bibinfo {volume} {81}},\ \bibinfo
  {pages} {155210} (\bibinfo {year} {2010})}\BibitemShut {NoStop}%
\bibitem [{\citenamefont {Koguchi}\ \emph {et~al.}(1992)\citenamefont
  {Koguchi}, \citenamefont {Kakibayashi}, \citenamefont {Yazawa}, \citenamefont
  {Hiruma},\ and\ \citenamefont {Katsuyama}}]{koguchi-JJAP92}%
  \BibitemOpen
  \bibfield  {author} {\bibinfo {author} {\bibfnamefont {M.}~\bibnamefont
  {Koguchi}}, \bibinfo {author} {\bibfnamefont {H.}~\bibnamefont
  {Kakibayashi}}, \bibinfo {author} {\bibfnamefont {M.}~\bibnamefont {Yazawa}},
  \bibinfo {author} {\bibfnamefont {K.}~\bibnamefont {Hiruma}}, \ and\ \bibinfo
  {author} {\bibfnamefont {T.}~\bibnamefont {Katsuyama}},\ }\href {\doibase
  10.1143/JJAP.31.2061} {\bibfield  {journal} {\bibinfo  {journal} {Japanese
  Journal of Applied Physics}\ }\textbf {\bibinfo {volume} {31}},\ \bibinfo
  {pages} {2061} (\bibinfo {year} {1992})}\BibitemShut {NoStop}%
\bibitem [{\citenamefont {Intronati}\ \emph {et~al.}(2013)\citenamefont
  {Intronati}, \citenamefont {Tamborenea}, \citenamefont {Weinmann},\ and\
  \citenamefont {Jalabert}}]{intronati-PRB2013}%
  \BibitemOpen
  \bibfield  {author} {\bibinfo {author} {\bibfnamefont {G.~A.}\ \bibnamefont
  {Intronati}}, \bibinfo {author} {\bibfnamefont {P.~I.}\ \bibnamefont
  {Tamborenea}}, \bibinfo {author} {\bibfnamefont {D.}~\bibnamefont
  {Weinmann}}, \ and\ \bibinfo {author} {\bibfnamefont {R.~A.}\ \bibnamefont
  {Jalabert}},\ }\href {\doibase 10.1103/PhysRevB.88.045303} {\bibfield
  {journal} {\bibinfo  {journal} {Phys. Rev. B}\ }\textbf {\bibinfo {volume}
  {88}},\ \bibinfo {pages} {045303} (\bibinfo {year} {2013})}\BibitemShut
  {NoStop}%
\bibitem [{\citenamefont {Intronati}(2013)}]{intronati_thesis}%
  \BibitemOpen
  \bibfield  {author} {\bibinfo {author} {\bibfnamefont {G.~A.}\ \bibnamefont
  {Intronati}},\ }\emph {\bibinfo {title} {Spin relaxation in doped
  semiconductors and semiconductor nanostructures}},\ \href
  {http://digital.bl.fcen.uba.ar/Download/Tesis/Tesis_5341_Intronati.pdf}
  {Ph.D. thesis},\ \bibinfo  {school} {Universidad de Buenos Aires} (\bibinfo
  {year} {2013})\BibitemShut {NoStop}%
\bibitem [{\citenamefont {Nozi\`eres}\ and\ \citenamefont
  {Lewiner}(1973)}]{noz-lew}%
  \BibitemOpen
  \bibfield  {author} {\bibinfo {author} {\bibfnamefont {P.}~\bibnamefont
  {Nozi\`eres}}\ and\ \bibinfo {author} {\bibfnamefont {C.}~\bibnamefont
  {Lewiner}},\ }\href {\doibase 10.1051/jphys:019730034010090100} {\bibfield
  {journal} {\bibinfo  {journal} {J. Phys. (Paris)}\ }\textbf {\bibinfo
  {volume} {34}},\ \bibinfo {pages} {901} (\bibinfo {year} {1973})}\BibitemShut
  {NoStop}%
\bibitem [{\citenamefont {Engel}\ \emph {et~al.}(2007)\citenamefont {Engel},
  \citenamefont {Rashba},\ and\ \citenamefont {Halperin}}]{eng-ras-hal}%
  \BibitemOpen
  \bibfield  {author} {\bibinfo {author} {\bibfnamefont {H.-A.}\ \bibnamefont
  {Engel}}, \bibinfo {author} {\bibfnamefont {E.~I.}\ \bibnamefont {Rashba}}, \
  and\ \bibinfo {author} {\bibfnamefont {B.~I.}\ \bibnamefont {Halperin}},\
  }\enquote {\bibinfo {title} {Theory of spin \uppercase{H}all effects in
  semiconductors},}\ in\ \href {\doibase 10.1002/9780470022184.hmm508} {\emph
  {\bibinfo {booktitle} {Handbook of Magnetism and Advanced Magnetic
  Materials}}}\ (\bibinfo  {publisher} {John Wiley and Sons, Ltd},\ \bibinfo
  {year} {2007})\BibitemShut {NoStop}%
\bibitem [{\citenamefont {Lew Yan~Voon}\ \emph {et~al.}(1996)\citenamefont {Lew
  Yan~Voon}, \citenamefont {Willatzen}, \citenamefont {Cardona},\ and\
  \citenamefont {Christensen}}]{lew-wil-car-chr}%
  \BibitemOpen
  \bibfield  {author} {\bibinfo {author} {\bibfnamefont {L.~C.}\ \bibnamefont
  {Lew Yan~Voon}}, \bibinfo {author} {\bibfnamefont {M.}~\bibnamefont
  {Willatzen}}, \bibinfo {author} {\bibfnamefont {M.}~\bibnamefont {Cardona}},
  \ and\ \bibinfo {author} {\bibfnamefont {N.~E.}\ \bibnamefont
  {Christensen}},\ }\href {\doibase 10.1103/PhysRevB.53.10703} {\bibfield
  {journal} {\bibinfo  {journal} {Phys. Rev. B}\ }\textbf {\bibinfo {volume}
  {53}},\ \bibinfo {pages} {10703} (\bibinfo {year} {1996})}\BibitemShut
  {NoStop}%
\bibitem [{\citenamefont {Meier}\ \emph {et~al.}(2007)\citenamefont {Meier},
  \citenamefont {Salis}, \citenamefont {Shorubalko}, \citenamefont {Gini},
  \citenamefont {Schon},\ and\ \citenamefont {Ensslin}}]{meier-NatPhy2007}%
  \BibitemOpen
  \bibfield  {author} {\bibinfo {author} {\bibfnamefont {L.}~\bibnamefont
  {Meier}}, \bibinfo {author} {\bibfnamefont {G.}~\bibnamefont {Salis}},
  \bibinfo {author} {\bibfnamefont {I.}~\bibnamefont {Shorubalko}}, \bibinfo
  {author} {\bibfnamefont {E.}~\bibnamefont {Gini}}, \bibinfo {author}
  {\bibfnamefont {S.}~\bibnamefont {Schon}}, \ and\ \bibinfo {author}
  {\bibfnamefont {K.}~\bibnamefont {Ensslin}},\ }\href {\doibase
  10.1038/nphys675} {\bibfield  {journal} {\bibinfo  {journal} {Nat. Phys.}\
  }\textbf {\bibinfo {volume} {3}},\ \bibinfo {pages} {650} (\bibinfo {year}
  {2007})}\BibitemShut {NoStop}%
\bibitem [{\citenamefont {Stefanowicz}\ \emph {et~al.}(2014)\citenamefont
  {Stefanowicz}, \citenamefont {Adhikari}, \citenamefont {Andrearczyk},
  \citenamefont {Faina}, \citenamefont {Sawicki}, \citenamefont {Majewski},
  \citenamefont {Dietl},\ and\ \citenamefont {Bonanni}}]{ste-2014}%
  \BibitemOpen
  \bibfield  {author} {\bibinfo {author} {\bibfnamefont {W.}~\bibnamefont
  {Stefanowicz}}, \bibinfo {author} {\bibfnamefont {R.}~\bibnamefont
  {Adhikari}}, \bibinfo {author} {\bibfnamefont {T.}~\bibnamefont
  {Andrearczyk}}, \bibinfo {author} {\bibfnamefont {B.}~\bibnamefont {Faina}},
  \bibinfo {author} {\bibfnamefont {M.}~\bibnamefont {Sawicki}}, \bibinfo
  {author} {\bibfnamefont {J.~A.}\ \bibnamefont {Majewski}}, \bibinfo {author}
  {\bibfnamefont {T.}~\bibnamefont {Dietl}}, \ and\ \bibinfo {author}
  {\bibfnamefont {A.}~\bibnamefont {Bonanni}},\ }\href {\doibase
  10.1103/PhysRevB.89.205201} {\bibfield  {journal} {\bibinfo  {journal} {Phys.
  Rev. B}\ }\textbf {\bibinfo {volume} {89}},\ \bibinfo {pages} {205201}
  (\bibinfo {year} {2014})}\BibitemShut {NoStop}%
\bibitem [{\citenamefont {Persson}\ \emph {et~al.}(2001)\citenamefont
  {Persson}, \citenamefont {Ferreira~da Silva}, \citenamefont {Ahuja},\ and\
  \citenamefont {Johansson}}]{per-fer-ahu-joh}%
  \BibitemOpen
  \bibfield  {author} {\bibinfo {author} {\bibfnamefont {C.}~\bibnamefont
  {Persson}}, \bibinfo {author} {\bibfnamefont {A.}~\bibnamefont {Ferreira~da
  Silva}}, \bibinfo {author} {\bibfnamefont {R.}~\bibnamefont {Ahuja}}, \ and\
  \bibinfo {author} {\bibfnamefont {B.}~\bibnamefont {Johansson}},\ }\href
  {\doibase https://doi.org/10.1016/S0022-0248(01)01470-1} {\bibfield
  {journal} {\bibinfo  {journal} {Journal of Crystal Growth}\ }\textbf
  {\bibinfo {volume} {231}},\ \bibinfo {pages} {397 } (\bibinfo {year}
  {2001})},\ \bibinfo {note} {proceedings of the International Specialist
  Meeting on Bulk Nitrides and Related Techniques}\BibitemShut {NoStop}%
\bibitem [{\citenamefont {Ferreira~da Silva}\ \emph {et~al.}(2001)\citenamefont
  {Ferreira~da Silva}, \citenamefont {Moys\'es~Ara\'ujo}, \citenamefont
  {Sernelius}, \citenamefont {Persson}, \citenamefont {Ahuja},\ and\
  \citenamefont {Johansson}}]{fer-ara}%
  \BibitemOpen
  \bibfield  {author} {\bibinfo {author} {\bibfnamefont {A.}~\bibnamefont
  {Ferreira~da Silva}}, \bibinfo {author} {\bibfnamefont {C.}~\bibnamefont
  {Moys\'es~Ara\'ujo}}, \bibinfo {author} {\bibfnamefont {B.~E.}\ \bibnamefont
  {Sernelius}}, \bibinfo {author} {\bibfnamefont {C.}~\bibnamefont {Persson}},
  \bibinfo {author} {\bibfnamefont {R.}~\bibnamefont {Ahuja}}, \ and\ \bibinfo
  {author} {\bibfnamefont {B.}~\bibnamefont {Johansson}},\ }\href
  {http://stacks.iop.org/0953-8984/13/i=40/a=303} {\bibfield  {journal}
  {\bibinfo  {journal} {J. Phys.: Condens. Matter}\ }\textbf {\bibinfo {volume}
  {13}},\ \bibinfo {pages} {8891} (\bibinfo {year} {2001})}\BibitemShut
  {NoStop}%
\bibitem [{\citenamefont {Rodina}\ \emph {et~al.}(2001)\citenamefont {Rodina},
  \citenamefont {Dietrich}, \citenamefont {G\"oldner}, \citenamefont {Eckey},
  \citenamefont {Hoffmann}, \citenamefont {Efros}, \citenamefont {Rosen},\ and\
  \citenamefont {Meyer}}]{Rodina2001}%
  \BibitemOpen
  \bibfield  {author} {\bibinfo {author} {\bibfnamefont {A.~V.}\ \bibnamefont
  {Rodina}}, \bibinfo {author} {\bibfnamefont {M.}~\bibnamefont {Dietrich}},
  \bibinfo {author} {\bibfnamefont {A.}~\bibnamefont {G\"oldner}}, \bibinfo
  {author} {\bibfnamefont {L.}~\bibnamefont {Eckey}}, \bibinfo {author}
  {\bibfnamefont {A.}~\bibnamefont {Hoffmann}}, \bibinfo {author}
  {\bibfnamefont {A.~L.}\ \bibnamefont {Efros}}, \bibinfo {author}
  {\bibfnamefont {M.}~\bibnamefont {Rosen}}, \ and\ \bibinfo {author}
  {\bibfnamefont {B.~K.}\ \bibnamefont {Meyer}},\ }\href {\doibase
  10.1103/PhysRevB.64.115204} {\bibfield  {journal} {\bibinfo  {journal} {Phys.
  Rev. B}\ }\textbf {\bibinfo {volume} {64}},\ \bibinfo {pages} {115204}
  (\bibinfo {year} {2001})}\BibitemShut {NoStop}%
\bibitem [{\citenamefont {Majlis}\ and\ \citenamefont
  {Anda}(1983)}]{maj-and_83}%
  \BibitemOpen
  \bibfield  {author} {\bibinfo {author} {\bibfnamefont {N.}~\bibnamefont
  {Majlis}}\ and\ \bibinfo {author} {\bibfnamefont {E.}~\bibnamefont {Anda}},\
  }\href@noop {} {\bibfield  {journal} {\bibinfo  {journal} {Solid State
  Comm.}\ }\textbf {\bibinfo {volume} {45}},\ \bibinfo {pages} {561} (\bibinfo
  {year} {1983})}\BibitemShut {NoStop}%
\bibitem [{\citenamefont {Figueira}\ \emph {et~al.}(1984)\citenamefont
  {Figueira}, \citenamefont {Makler},\ and\ \citenamefont
  {Anda}}]{fig-mak-and_84}%
  \BibitemOpen
  \bibfield  {author} {\bibinfo {author} {\bibfnamefont {M.}~\bibnamefont
  {Figueira}}, \bibinfo {author} {\bibfnamefont {S.}~\bibnamefont {Makler}}, \
  and\ \bibinfo {author} {\bibfnamefont {E.}~\bibnamefont {Anda}},\ }\href@noop
  {} {\bibfield  {journal} {\bibinfo  {journal} {J. Phys. C: Solid State
  Phys.}\ }\textbf {\bibinfo {volume} {17}},\ \bibinfo {pages} {623} (\bibinfo
  {year} {1984})}\BibitemShut {NoStop}%
\bibitem [{\citenamefont {Roberts}\ and\ \citenamefont
  {Ursell}(1960)}]{roberts-ursell-1960}%
  \BibitemOpen
  \bibfield  {author} {\bibinfo {author} {\bibfnamefont {P.~H.}\ \bibnamefont
  {Roberts}}\ and\ \bibinfo {author} {\bibfnamefont {H.~D.}\ \bibnamefont
  {Ursell}},\ }\href {\doibase 10.1098/rsta.1960.0008} {\bibfield  {journal}
  {\bibinfo  {journal} {Phil. Trans. R. Soc. Lond. A}\ }\textbf {\bibinfo
  {volume} {252}},\ \bibinfo {pages} {317} (\bibinfo {year}
  {1960})}\BibitemShut {NoStop}%
\bibitem [{\citenamefont {Vollhardt}\ and\ \citenamefont
  {W\"olfle}(1980)}]{vol-woe}%
  \BibitemOpen
  \bibfield  {author} {\bibinfo {author} {\bibfnamefont {D.}~\bibnamefont
  {Vollhardt}}\ and\ \bibinfo {author} {\bibfnamefont {P.}~\bibnamefont
  {W\"olfle}},\ }\href {\doibase 10.1103/PhysRevB.22.4666} {\bibfield
  {journal} {\bibinfo  {journal} {Phys. Rev. B}\ }\textbf {\bibinfo {volume}
  {22}},\ \bibinfo {pages} {4666} (\bibinfo {year} {1980})}\BibitemShut
  {NoStop}%
\bibitem [{\citenamefont {Vollhardt}\ and\ \citenamefont
  {W\"olfle}(1992)}]{rev-vol-woe}%
  \BibitemOpen
  \bibfield  {author} {\bibinfo {author} {\bibfnamefont {D.}~\bibnamefont
  {Vollhardt}}\ and\ \bibinfo {author} {\bibfnamefont {P.}~\bibnamefont
  {W\"olfle}},\ }\enquote {\bibinfo {title} {Self-consistent theory of
  \uppercase{A}nderson localization},}\ in\ \href
  {https://www.elsevier.com/books/electronic-phase-transitions/kopaev/978-0-444-88885-3}
  {\emph {\bibinfo {booktitle} {Electronic phase transitions}}},\ \bibinfo
  {editor} {edited by\ \bibinfo {editor} {\bibfnamefont {W.}~\bibnamefont
  {Hanke}}\ and\ \bibinfo {editor} {\bibfnamefont {Y.~V.}\ \bibnamefont
  {Kopaev}}}\ (\bibinfo  {publisher} {North-Holland},\ \bibinfo {address}
  {Amsterdam},\ \bibinfo {year} {1992})\ p.~\bibinfo {pages} {1}\BibitemShut
  {NoStop}%
\bibitem [{\citenamefont {Kroha}\ \emph {et~al.}(1990)\citenamefont {Kroha},
  \citenamefont {Kopp},\ and\ \citenamefont {W\"olfle}}]{kkw90}%
  \BibitemOpen
  \bibfield  {author} {\bibinfo {author} {\bibfnamefont {J.}~\bibnamefont
  {Kroha}}, \bibinfo {author} {\bibfnamefont {T.}~\bibnamefont {Kopp}}, \ and\
  \bibinfo {author} {\bibfnamefont {P.}~\bibnamefont {W\"olfle}},\ }\href
  {\doibase 10.1103/PhysRevB.41.888} {\bibfield  {journal} {\bibinfo  {journal}
  {Phys. Rev. B}\ }\textbf {\bibinfo {volume} {41}},\ \bibinfo {pages} {888}
  (\bibinfo {year} {1990})}\BibitemShut {NoStop}%
\bibitem [{\citenamefont {Elyutin}(1981)}]{Elyutin}%
  \BibitemOpen
  \bibfield  {author} {\bibinfo {author} {\bibfnamefont {P.}~\bibnamefont
  {Elyutin}},\ }\href
  {http://iopscience.iop.org/article/10.1088/0022-3719/14/10/011/pdf}
  {\bibfield  {journal} {\bibinfo  {journal} {J. Phys. C: Solid State Phys.}\
  }\textbf {\bibinfo {volume} {14}},\ \bibinfo {pages} {1435} (\bibinfo {year}
  {1981})}\BibitemShut {NoStop}%
\bibitem [{\citenamefont {Ferreira~da Silva}\ and\ \citenamefont
  {Persson}(2002)}]{fer-per}%
  \BibitemOpen
  \bibfield  {author} {\bibinfo {author} {\bibfnamefont {A.}~\bibnamefont
  {Ferreira~da Silva}}\ and\ \bibinfo {author} {\bibfnamefont {C.}~\bibnamefont
  {Persson}},\ }\href {\doibase 10.1063/1.1499202} {\bibfield  {journal}
  {\bibinfo  {journal} {J.\ Appl.\ Phys.}\ }\textbf {\bibinfo {volume} {92}},\
  \bibinfo {pages} {2550} (\bibinfo {year} {2002})}\BibitemShut {NoStop}%
\bibitem [{\citenamefont {Yin}\ \emph {et~al.}(2010)\citenamefont {Yin},
  \citenamefont {Shen}, \citenamefont {Zhang}, \citenamefont {Xu},
  \citenamefont {Tang}, \citenamefont {Cen}, \citenamefont {Wang},
  \citenamefont {Chen},\ and\ \citenamefont {Yu}}]{yin-APL2010}%
  \BibitemOpen
  \bibfield  {author} {\bibinfo {author} {\bibfnamefont {C.}~\bibnamefont
  {Yin}}, \bibinfo {author} {\bibfnamefont {B.}~\bibnamefont {Shen}}, \bibinfo
  {author} {\bibfnamefont {Q.}~\bibnamefont {Zhang}}, \bibinfo {author}
  {\bibfnamefont {F.}~\bibnamefont {Xu}}, \bibinfo {author} {\bibfnamefont
  {N.}~\bibnamefont {Tang}}, \bibinfo {author} {\bibfnamefont {L.}~\bibnamefont
  {Cen}}, \bibinfo {author} {\bibfnamefont {X.}~\bibnamefont {Wang}}, \bibinfo
  {author} {\bibfnamefont {Y.}~\bibnamefont {Chen}}, \ and\ \bibinfo {author}
  {\bibfnamefont {J.}~\bibnamefont {Yu}},\ }\href {\doibase 10.1063/1.3511768}
  {\bibfield  {journal} {\bibinfo  {journal} {Appl. Phys. Lett.}\ }\textbf
  {\bibinfo {volume} {97}},\ \bibinfo {pages} {181904} (\bibinfo {year}
  {2010})}\BibitemShut {NoStop}%
\bibitem [{\citenamefont {Andrearczyk}\ \emph {et~al.}(2005)\citenamefont
  {Andrearczyk}, \citenamefont {Jaroszy\ifmmode~\acute{n}\else \'{n}\fi{}ski},
  \citenamefont {Grabecki}, \citenamefont {Dietl}, \citenamefont {Fukumura},\
  and\ \citenamefont {Kawasaki}}]{and-jar-gra}%
  \BibitemOpen
  \bibfield  {author} {\bibinfo {author} {\bibfnamefont {T.}~\bibnamefont
  {Andrearczyk}}, \bibinfo {author} {\bibfnamefont {J.}~\bibnamefont
  {Jaroszy\ifmmode~\acute{n}\else \'{n}\fi{}ski}}, \bibinfo {author}
  {\bibfnamefont {G.}~\bibnamefont {Grabecki}}, \bibinfo {author}
  {\bibfnamefont {T.}~\bibnamefont {Dietl}}, \bibinfo {author} {\bibfnamefont
  {T.}~\bibnamefont {Fukumura}}, \ and\ \bibinfo {author} {\bibfnamefont
  {M.}~\bibnamefont {Kawasaki}},\ }\href {\doibase 10.1103/PhysRevB.72.121309}
  {\bibfield  {journal} {\bibinfo  {journal} {Phys. Rev. B}\ }\textbf {\bibinfo
  {volume} {72}},\ \bibinfo {pages} {121309} (\bibinfo {year}
  {2005})}\BibitemShut {NoStop}%
\bibitem [{\citenamefont {Majewski}(2005)}]{maj-2005-APP}%
  \BibitemOpen
  \bibfield  {author} {\bibinfo {author} {\bibfnamefont {J.~A.}\ \bibnamefont
  {Majewski}},\ }\href {\doibase 10.12693/APhysPolA.108.777} {\bibfield
  {journal} {\bibinfo  {journal} {Acta Phys.\ Pol.\ A}\ }\textbf {\bibinfo
  {volume} {108}},\ \bibinfo {pages} {777} (\bibinfo {year}
  {2005})}\BibitemShut {NoStop}%
\bibitem [{\citenamefont {Majewski}\ and\ \citenamefont
  {Vogl}(2005)}]{maj-2005}%
  \BibitemOpen
  \bibfield  {author} {\bibinfo {author} {\bibfnamefont {J.~A.}\ \bibnamefont
  {Majewski}}\ and\ \bibinfo {author} {\bibfnamefont {P.}~\bibnamefont
  {Vogl}},\ }in\ \href@noop {} {\emph {\bibinfo {booktitle} {Physics of
  Semiconductors: 27th International Conference on the Physics of
  Semiconductors}}},\ \bibinfo {editor} {edited by\ \bibinfo {editor}
  {\bibfnamefont {J.}~\bibnamefont {Men\'endez}}\ and\ \bibinfo {editor}
  {\bibfnamefont {C.~G.}\ \bibnamefont {Van~de Walle}}}\ (\bibinfo  {publisher}
  {American Institute of Physics},\ \bibinfo {address} {New York},\ \bibinfo
  {year} {2005})\ p.\ \bibinfo {pages} {1403}\BibitemShut {NoStop}%
\bibitem [{\citenamefont {Scher\"ubl}\ \emph {et~al.}(2016)\citenamefont
  {Scher\"ubl}, \citenamefont {F\"ul\"op}, \citenamefont {Madsen},
  \citenamefont {Nyg\aa{}rd},\ and\ \citenamefont
  {Csonka}}]{sch-zol-fue-mad-nyg-cso}%
  \BibitemOpen
  \bibfield  {author} {\bibinfo {author} {\bibfnamefont {Z.}~\bibnamefont
  {Scher\"ubl}}, \bibinfo {author} {\bibfnamefont {G.}~\bibnamefont
  {F\"ul\"op}}, \bibinfo {author} {\bibfnamefont {M.~H.}\ \bibnamefont
  {Madsen}}, \bibinfo {author} {\bibfnamefont {J.}~\bibnamefont {Nyg\aa{}rd}},
  \ and\ \bibinfo {author} {\bibfnamefont {S.}~\bibnamefont {Csonka}},\ }\href
  {\doibase 10.1103/PhysRevB.94.035444} {\bibfield  {journal} {\bibinfo
  {journal} {Phys. Rev. B}\ }\textbf {\bibinfo {volume} {94}},\ \bibinfo
  {pages} {035444} (\bibinfo {year} {2016})}\BibitemShut {NoStop}%
\bibitem [{\citenamefont {Liang}\ and\ \citenamefont {Gao}(2012)}]{lia-xua}%
  \BibitemOpen
  \bibfield  {author} {\bibinfo {author} {\bibfnamefont {D.}~\bibnamefont
  {Liang}}\ and\ \bibinfo {author} {\bibfnamefont {X.~P.}\ \bibnamefont
  {Gao}},\ }\href {\doibase 10.1021/nl301325h} {\bibfield  {journal} {\bibinfo
  {journal} {Nano Letters}\ }\textbf {\bibinfo {volume} {12}},\ \bibinfo
  {pages} {3263} (\bibinfo {year} {2012})}\BibitemShut {NoStop}%
\bibitem [{\citenamefont {Wolos}\ \emph
  {et~al.}(2011{\natexlab{b}})\citenamefont {Wolos}, \citenamefont
  {Wilamowski}, \citenamefont {Skierbiszewski}, \citenamefont {Drabinska},
  \citenamefont {Lucznik}, \citenamefont {Grzegory},\ and\ \citenamefont
  {Porowski}}]{wol-2011}%
  \BibitemOpen
  \bibfield  {author} {\bibinfo {author} {\bibfnamefont {A.}~\bibnamefont
  {Wolos}}, \bibinfo {author} {\bibfnamefont {Z.}~\bibnamefont {Wilamowski}},
  \bibinfo {author} {\bibfnamefont {C.}~\bibnamefont {Skierbiszewski}},
  \bibinfo {author} {\bibfnamefont {A.}~\bibnamefont {Drabinska}}, \bibinfo
  {author} {\bibfnamefont {B.}~\bibnamefont {Lucznik}}, \bibinfo {author}
  {\bibfnamefont {I.}~\bibnamefont {Grzegory}}, \ and\ \bibinfo {author}
  {\bibfnamefont {S.}~\bibnamefont {Porowski}},\ }\href {\doibase
  https://doi.org/10.1016/j.physb.2011.03.060} {\bibfield  {journal} {\bibinfo
  {journal} {Physica B: Condensed Matter}\ }\textbf {\bibinfo {volume} {406}},\
  \bibinfo {pages} {2548 } (\bibinfo {year} {2011}{\natexlab{b}})}\BibitemShut
  {NoStop}%
\end{thebibliography}%

\end{document}